\documentclass[]{aastex631}

\usepackage{amsmath}

\usepackage{ulem}
\usepackage{xcolor}
\definecolor{blazeorange}{rgb}{1.0, 0.4, 0.0}
\definecolor{seagreen}{rgb}{0.18, 0.55, 0.34}
\definecolor{rufous}{rgb}{0.66, 0.11, 0.03}
\definecolor{royalfuchsia}{rgb}{0.79, 0.17, 0.57}
\definecolor{scarlet}{rgb}{1.0, 0.13, 0.0}
\definecolor{rose}{rgb}{1.0, 0.13, 0.5}
\definecolor{royalpurple}{rgb}{0.47, 0.32, 0.66}
\definecolor{darkblue}{rgb}{0, 0, 0.66}

\begin{document}

\title{The Cosmic Evolution of Fast Radio Bursts Inferred from the CHIME/FRB Baseband Catalog 1}

\correspondingauthor{Om Gupta}
\email{omgupta@utexas.edu}

\author[0000-0001-8470-7289]{Om Gupta}
\affiliation{Department of Astronomy, The University of Texas at Austin, Austin, TX 78712, USA}

\author[0000-0001-7833-1043]{Paz Beniamini}
\affiliation{Department of Natural Sciences, The Open University of Israel, P.O Box 808, Ra’anana 4353701, Israel}
\affiliation{Astrophysics Research Center of the Open University (ARCO), The Open University of Israel, P.O Box 808, Ra’anana 4353701, Israel}
\affiliation{Department of Physics, The George Washington University, 725 21st Street NW, Washington, DC 20052, USA}

\author[0009-0003-8955-7402]{Pawan Kumar}
\affiliation{Department of Astronomy, The University of Texas at Austin, Austin, TX 78712, USA}

\author[0000-0001-8519-1130]{Steven L. Finkelstein}
\affiliation{Department of Astronomy, The University of Texas at Austin, Austin, TX 78712, USA}



\newcommand{\energyunit}{erg Hz$^{-1}$}
\newcommand{\dmunit}{pc cm$^{-3}$}
\newcommand{\echar}{$E_{\rm char, \nu}\;$}
\newcommand{\volrate}{Mpc$^{-3}$ yr$^{-1}$}
\newcommand{\gvolrate}{Gpc$^{-3}$ yr$^{-1}$}

\begin{abstract}

Redshift and luminosity distributions are essential for understanding the cosmic evolution of extragalactic objects and phenomena, such as galaxies, gamma-ray bursts, and fast radio bursts (FRBs). For FRBs, these distributions are primarily estimated using the fluence and the Dispersion Measure (DM). Calibrating their joint distribution has been challenging due to a lack of accurate fluences in the intensity data of the CHIME/FRB survey. Using the baseband update of CHIME/FRB Catalog 1, we calibrate the 2D fluence-DM distribution for the first time. We find the energy distribution is described well by a Schechter function with power-law slope of $-1.94^{+0.14}_{-0.12}$.
Testing two types of redshift evolution models suggests a likely combination of young and old formation channels. $31^{+31}_{-21}$\% of FRB sources may track star formation, or correspondingly, FRB sources may have delay times of $1.94^{+1.54}_{-1.31}$ Gyr. A pure star formation tracking population is excluded by only one model at $> 2\sigma$ confidence.
An updated cosmic star formation rate density evolution up to redshift 14 is constrained by compiling results from several JWST studies. The furthest FRB detection with planned radio facilities is expected to be at $z \approx 5$.
A radio telescope operating at 200 MHz with a system-equivalent flux density of $\leq 0.07$ Jy (equivalent to a detection threshold of 1 mJy ms) and instantaneous sky coverage of $\gtrsim 400$ square degrees should be able to detect $630^{+730}_{-485}$ FRBs year$^{-1}$ at $z \gtrsim 6$ and $53^{+83}_{-43}$ FRBs year$^{-1}$ at $z\gtrsim 8$, which is sufficient to differentiate between reionization histories.

\end{abstract}

\keywords{Radio transient sources (2008) --- Radio bursts (1339) --- Extragalactic radio sources (508) --- Luminosity function (942)}


\section{Introduction} \label{sec:intro}

Fast Radio Bursts (FRBs) are radio signals with time durations ranging from microseconds to tens of milliseconds, detected with spectral irradiances on the order of Janskys, most commonly having an extragalactic source.
The release of the First CHIME/FRB Catalog \citep{CHIME2021} from the Canadian Hydrogen Intensity Mapping Experiment (CHIME) has been a huge milestone in the field, as the number of FRBs jumped from less than 100 detected bursts to around 600. There are now more than 800 FRB sources detected, out of which 67 have produced repeat bursts, as determined by a compiled catalog of bursts available on Blinkverse \citep{Xu2023_Blinkverse}. 

The extremely high brightness temperatures inferred from FRBs with a nominal value of $T_b \sim 10^{36}$ K, much higher than those observed in radio pulsars of around $\sim 10^{25} - 10^{30}$ K \citep{Zhang2022review}, indicate that FRBs are also powered by coherent emission processes. It is widely believed that FRBs may most commonly be emitted from the environment around magnetars (see e.g., \citealt{Popov2010, Kumar2017,KumarBosnjak2020}), and this belief was bolstered by the observation of FRB-like bursts from a known Galactic magnetar, SGR 1935+2154 \citep{CHIME2020, Bochenek2020, Kirsten2021, Giri2023}.
The repeating FRB 20200120E has been localized to a globular cluster of the nearby galaxy M81 \citep{Bhardwaj2021, Kirsten2022}, which is devoid of any recent star formation. This progenitor could be the result of either accretion-induced collapse of a white dwarf, compact binary merger, or spin-up of an accretion-recycled neutron star \citep{Kirsten2022, Li2023}. A wide variety of models to explain FRB sources have been proposed in the literature (e.g., \citealt{Cordes2016, Katz2017, Dai2016, Cooper2023}).

The activity periods of magnetars are expected to be very short compared to cosmological timescales, due to the decay of their strong magnetic fields. Even ultra-long period magnetars may be able to retain their magnetic fields only up to $\sim$Myr \citep{Beniamini2020, Beniamini2023}.
As such, it is expected that the formation rate of the FRB source population, if it is dominated by SGR-like sources formed through core-collapse supernovae, may effectively track the cosmic star formation rate (SFR). However, the globular cluster FRB relates a fraction of the FRB population to late-stage evolutionary scenarios, which predict an FRB rate that increases with cosmic time \citep{James2022a}. The cosmic evolution of these late-stage scenarios may be modelled either by the stellar mass density (SMD) or by choosing a certain delay time model. More realistically, the FRB population may have a cosmic evolutionary track created by several FRB progenitor formation channels that themselves evolve differently with cosmic time. Constraining the cosmic evolution of FRBs has generated a lot of interest in recent years as the number of observed FRBs has increased. It is also synergistic to determining the energy function of FRBs. 

A number of studies have tried to constrain the FRB redshift evolution, with some of them using data from Parkes and ASKAP, such as \citet{Zhang2021} who find that the redshift distribution cannot be constrained, and \citet{James2022a} who find that FRBs track the cosmic SFR. Most of the studies use FRB data from the CHIME telescope which performs a blind survey and has a large field-of-view of the sky.
The CHIME/FRB Catalog 1 \citep{CHIME2021} provides the largest uniform sample of FRBs, given that all bursts in the catalog have been detected by the same telescope in a single survey and suffers from the same biases (against which significant effort has been made to correct). It has 536 bursts in total, including 474 bursts from (apparently) non-repeating sources, and 62 bursts from 18 repeating sources. As such, this sample presents the clearest opportunity to model FRB energy and redshift distributions, as has been attempted in several previous publications (e.g., \citealt{Shin2023, Hashimoto2022, Bhattacharyya2023, Tang2023}). Some studies find that FRBs evolve with SMD \citep{Hashimoto2022, Li2023_zou} and some studies rule out SFR-tracking models \citep{Lin2024, Qiang2022, Zhang2022}, finding that models which include delays with respect to SFR or those having a redshift-parameterized SFR are better fits to the data. Using the CHIME injections pipeline to determine the energy and redshift distributions, \citet{Shin2023} parameterize the SFR with an exponent, finding a median value which is consistent with SFR but with large confidence intervals that leave it poorly constrained.

Building on the efforts of previous studies, we use CHIME/FRB Baseband data \citep{BasebandCatalog1_2023} to constrain how FRBs evolve at low redshifts. We use two models providing complementary information to better understand the underlying source population. The first is a hybrid SFR-SMD model designed to infer the mixture of early- and late-type evolutionary scenarios, while the second is a constant delay time model which provides an estimate of the average delay time of FRB sources with respect to star formation. Our aim is also to constrain an updated cosmic star formation rate density in the era of JWST. With the calibration of low-redshift FRB evolution, we use the cosmic star formation history inferred from published JWST data 
to include forecasts of the number of FRBs observable by future radio telescopes, such as CHORD \citep{Vanderlinde2019}, DSA-2000 \citep{Hallinan2019}, BURSTT \citep{Lin2022_burstt} and the Square Kilometer Array (SKA; \citealt{Braun2019_SKA1}). We also estimate the specifications required for a radio telescope that should be able to probe the Epoch of Reionization. These exercises help us understand whether FRBs can be used as probes of the high-redshift Universe with these next-generation telescopes to constrain cosmology and the epochs of hydrogen and helium reionization, and what the possibilities are to improve the observational redshift horizon.

Section \ref{sec:sfrd_evol} describes the computation of the star formation rate density between redshifts 0 and 14 (uninterested readers may skip this section altogether). Section \ref{sec:cat1_des} describes the data and its selection process. Section \ref{sec:theory_model} outlines the theoretical and statistical model powering the analysis. Section \ref{sec:results} presents the results of the analysis and makes predictions for future telescopes. Section \ref{sec:discussion} probes the analysis and the pipeline, but additionally discusses the broader consequences of this work, relating it to recent results about FRB populations and host galaxies. Finally, the important conclusions of this study are presented in Section \ref{sec:conclusions}.

\section{Evolution of the Star Formation Rate Density} \label{sec:sfrd_evol}

The FRB progenitor population evolution is directly related to the underlying stellar population density. For progenitor scenarios involving neutron stars or black holes, it is either the death of massive stars or binary mergers, among a few other channels, that creates them. A crucial indicator of this process is the star formation rate density (SFRD), denoted by $\dot{m}_{\ast}$, and it is expected that FRB progenitors will scale as some function of the SFRD. 

Deployment of JWST has increased human capability beyond Hubble's limit to observe star formation and galaxy evolution at and maybe even beyond a redshift of 14. A number of early-JWST surveys have already provided estimates of the observed galaxy densities in their survey volumes, which has opened up the possibility of estimating the SFRD at high redshifts (e.g. \citealt{Finkelstein2024, Robertson2024}). 
Very bright galaxies are rare and hence, large survey volumes are required to find them. On the other hand, faint galaxies may be numerous, but they need deeper (long exposure) observations. In the coming years, more data from JWST will be able to add sufficient number of galaxies, at both the bright and faint end, to estimate SFRDs more accurately at high-z.

\subsection{Constraining the UV Luminosity Function at High Redshift}
At low redshifts, UV and IR surveys probing star formation provide information across a wide range of stellar masses \citep{Madau2014}. However, for high redshift galaxies, the availability of only rest-frame UV light provides information just about the massive short-lived stellar population. The predictions we make would provide a lower limit to the FRB rate, so long as the fraction of sources formed via each progenitor channel does not change. In reality, the initial mass function (IMF) shifts to higher stellar masses at high redshifts, thereby forming more possible neutron star FRB progenitors from the core-collapse of massive stars. Additionally, channels requiring large delay times will cease to operate at sufficiently high redshift. A complex evolution of FRB sources incorporating a variety of undetermined factors is unfeasible presently.

The non-ionizing UV luminosity density can be estimated by integrating over the first moment of the UV luminosity function (UVLF) - the volume density of galaxies with a given non-ionizing UV luminosity. A functional form of the UVLF at high-z is obtained by fitting observational data of deep photometric surveys that count the number of galaxies in the estimated survey volume in several UV-magnitude bins. Historically, the functional form of the UVLF has been taken to be a Schechter function, but recently \cite{Finkelstein2022} have shown that a double power-law (DPL) is a more accurate fit to the data. DPL is defined in terms of absolute magnitudes as,
\begin{equation}
    \phi(M) = \phi_{\ast} (10^{0.4(\eta+1)(M-M_{\ast})} + 10^{0.4(\beta+1)(M-M_{\ast})})^{-1}
    \label{eq:dpl}
\end{equation}
where, $\eta$ is the faint-end slope, $\beta$ is the bright-end slope, $M_{\ast}$ is the characteristic magnitude at the junction of the two power-laws, and $\phi_{\ast}$ is the characteristic volume number density per magnitude at the knee.

\begin{figure*}
    \gridline{\fig{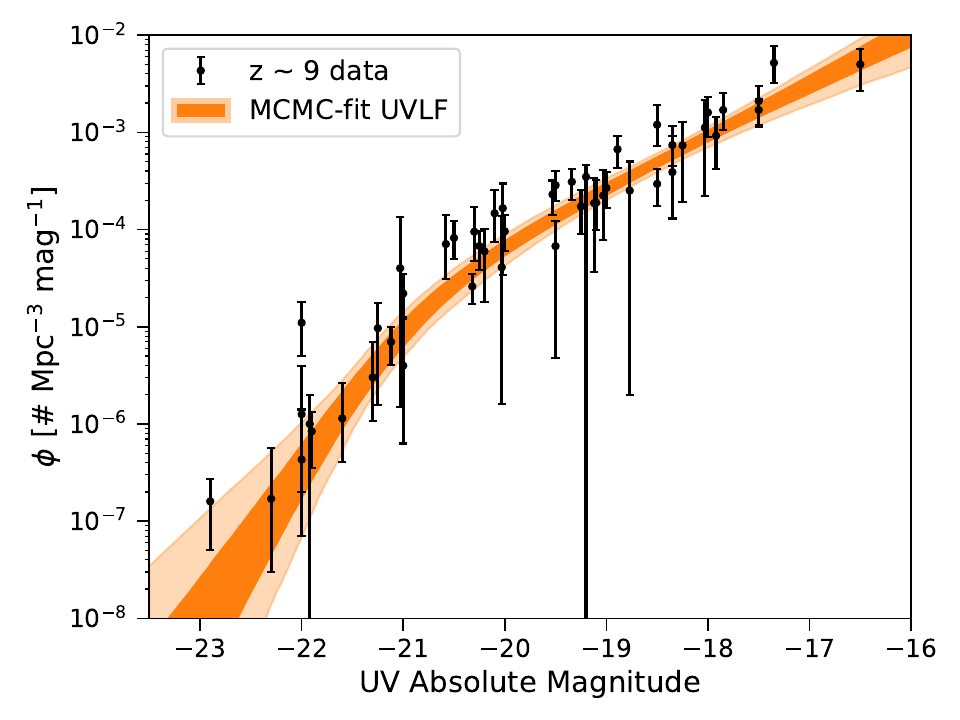}{0.49\textwidth}{(a)}
          \fig{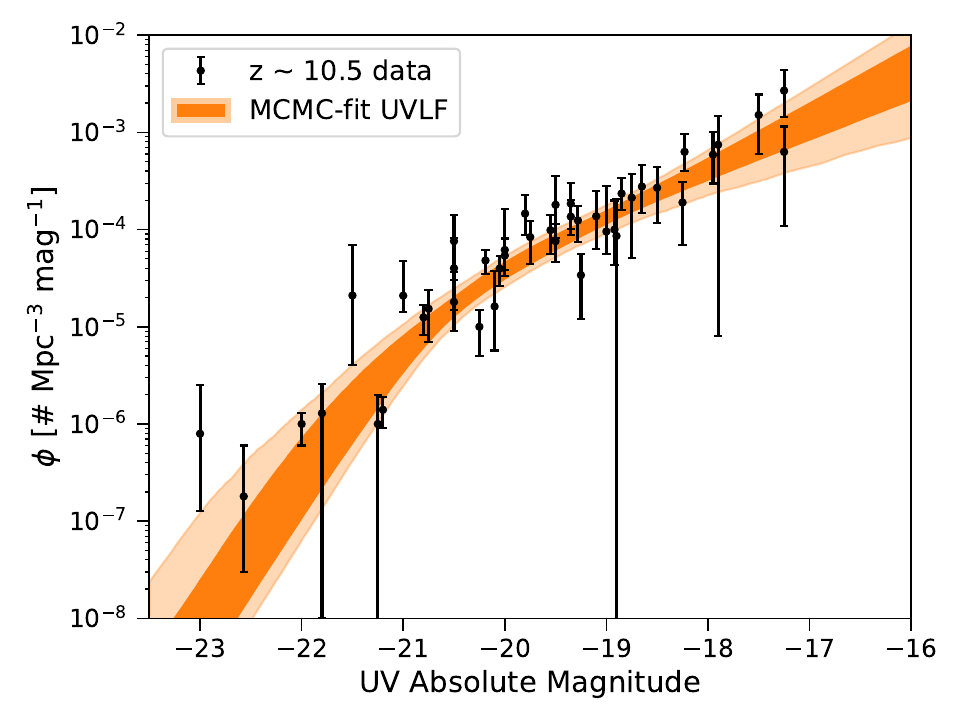}{0.49\textwidth}{(b)}}
    \gridline{\fig{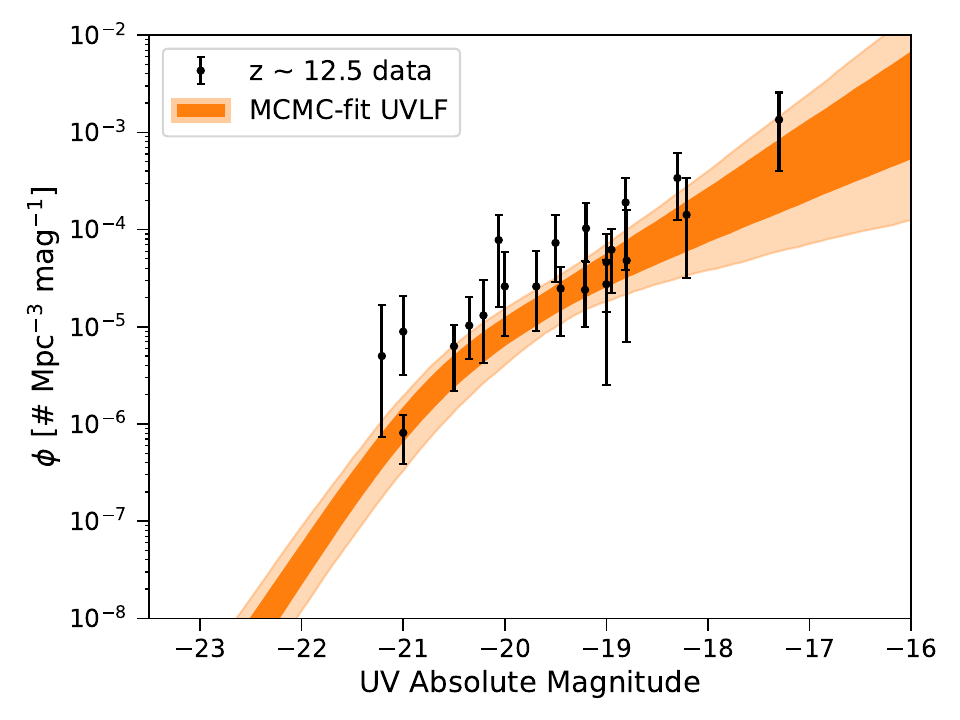}{0.49\textwidth}{(c)}}
    \caption{The rest-frame non-ionizing UV luminosity functions from redshifts z=9-14. a) $8.7 \leq z \leq 9$, b) $10 \leq z \leq 11$, and c) $11.2 \leq z \leq 14$. Data points are taken from the following references: \citet{Mclure2013, McLeod2016, Oesch2018, Morishita2018, Stefanon2019, Bowler2020, Bouwens2021, Naidu2022, Bouwens2023, Donnan2023a, Donnan2023b, Harikane2023, Leung2023, Perez-Gonzalez2023, Castellano2023, Finkelstein2023, McLeod2024, Adams2024, Franco2024, Casey2024, Willott2024, Chemerynska2024}.}
    \label{fig:lumfuncs}
\end{figure*}

We obtain luminosity function data at high redshifts from JWST, HST and other deep photometric and spectroscopic surveys. The data are divided into three redshift bins centered at $\sim$ 9, 10.5, and 12.5, with 49, 43, and 21 data points respectively, which are plotted in Figure \ref{fig:lumfuncs}. We use the DPL to describe the UVLF in each redshift bin, and constrain its parameters using a Markov Chain Monte Carlo (MCMC) ensemble sampling of the parameter space. This analysis to obtain the posterior distributions of the DPL parameters is done using the \textit{emcee} Python package \citep{Foreman-Mackey2013}. The initial parameter values provided to the MCMC sampler are obtained by maximum-likelihood estimation of the DPL parameters on the data in each redshift bin.

As there is sufficient data to constrain all parameters for z $\sim$ 9, we choose uniform priors for all DPL parameters. In higher redshift bins, the sparsity of the data impacts the determination of the bright-end DPL slope $\beta$. For z $\sim$ 10.5, we implement a Gaussian prior on $\beta$. For the z $\sim12.5$ bin, the data are very sparse, and as such we implement Gaussian priors on both $\beta$ and $M_*$. While taking Gaussian priors, we assume no evolution of these parameters between the redshift bins, and fix the median of the Gaussian prior to the median of the posterior distribution of the same parameter in the lower redshift bin. The exact priors used for all parameters are listed in Table \ref{tab:uvlf_data}.

The MCMC chain is run with 32 walkers for 50,000 steps, and the first 2000 steps are discarded. Non-Gaussian data errors are accounted for in the likelihood function; if the MCMC draw is lower than the observed mean value, then the lower error is assigned, and vice versa. To measure the performance of the algorithm, it is checked that the chains are longer than 75 times the integrated autocorrelation length for each parameter \citep{Foreman-Mackey2013}. 
The chains do not converge too slowly to be concerning, and the constraints on DPL parameters for all 3 redshift bins are detailed in Table \ref{tab:uvlf_data}. The corresponding DPLs are plotted in Figure \ref{fig:lumfuncs}.

\begin{deluxetable*}{cccccccccc} \label{tab:uvlf_data}
\tablecaption{Double power-law parameter constraints for the galaxy UV luminosity function.}
\tablehead{\colhead{z} & \multicolumn{2}{c}{$log_{10}\phi_{\ast}$ (Mpc$^{-3}$)} & \multicolumn{2}{c}{$M_{\ast}$ (mag)} & \multicolumn{2}{c}{$\eta$}  & \multicolumn{2}{c}{$\beta$} \\ & \colhead{Prior} & \colhead{Posterior} & \colhead{Prior} & \colhead{Posterior} & \colhead{Prior} & \colhead{Posterior} & \colhead{Prior} & \colhead{Posterior}}
\startdata
9 & [-10, -1] & $-4.55^{+0.43}_{-0.34}$ & [-24, -15] & $-20.81^{+0.53}_{-0.38}$ & [-4, -1] & $-2.34^{+0.17}_{-0.13}$ & [-8, -3] & $-5.02^{+0.84}_{-1.19}$\\
10.5 & [-10, -1] & $-4.80^{+0.35}_{-0.53}$ & [-24, -15] & $-20.84^{+0.39}_{-0.71}$ & [-4, -1] & $-2.25^{+0.25}_{-0.20}$ & $\mathcal{N}(\beta_{z=9}, 0.1)$ & $-5.01^{+0.10}_{-0.10}$\\
12.5 & [-10, -1] & $-5.48^{+0.19}_{-0.22}$ & $\mathcal{N}(M_{\ast, z=10.5}, 0.1)$ & $-20.83^{+0.10}_{-0.10}$ & [-4, -1] & $-2.47^{+0.31}_{-0.32}$ & $\mathcal{N}(\beta_{z=10.5}, 0.1)$ & $-5.01^{+0.10}_{-0.10}$\\
\enddata
\tablecomments{$\mathcal{N}$ refers to a normal distribution, with mean and standard deviation noted in the parentheses.}
\end{deluxetable*}

\subsection{Cosmic Evolution of the Star-formation Rate Density}

The massive SFRD can now be quantified based on galaxy UV luminosities between $0<z<14$, by combining JWST data with low-redshift SFRD information readily available (e.g., in \citealt{Madau2014, Finkelstein2022}). It is important to note that estimating SFRD requires several assumptions, which simplify the process but are known to introduce substantial errors. Moreover, this problem exacerbates at higher redshifts where large areas of the sky cannot be surveyed, and survey completeness estimates may be error-prone.

At low-redshifts between $0<z<4$, we directly use SFRD data from various surveys reported in \citet{Madau2014}. We then use the DPL parameters describing the UVLF in \citet{Finkelstein2022} to calculate the specific UV luminosity density $\rho_{UV}$ between $3 \leq z \leq 9$, assuming their parameters have Gaussian errors with the larger reported error as the standard deviation. We use our estimate of the DPL parameters to calculate $\rho_{UV}$ between $9 \leq z \leq 14$. Because dust attenuation is high in the UV, we also need to correct for dust while calculating $\rho_{UV}$, with the process outlined and discussed in Appendix \ref{sec:dustcorr}. It is assumed that the $z\geq 9$ data are unaffected by dust, and hence, we do not correct for it.

To calculate the specific UV luminosity density, we use the following relation,
\begin{equation}
    \rho_{\rm UV} = 10^{-0.4(\mathcal{M} + \mu(z) + 48.6)} \frac{4\pi d_L^2(z)}{1+z}\;{\rm erg\;s^{-1}\;Mpc^{-3}\;Hz^{-1}},
\end{equation}
where, $d_L(z)$ is the luminosity distance (in cm) at the observed redshift $z$, $\mu(z) = 5\;log_{10}\Big( \frac{1}{1+z}\frac{d_L}{10\;{\rm pc}} \Big)$ is the distance modulus, $48.6$ is the zero-point magnitude for the AB magnitude system when using cgs units for the specific flux density, and,
\begin{equation}
    \mathcal{M} = -2.5 \; log_{10} \Big( \int_{M_1}^{M_2} 10^{-0.4M} \phi(M) dM \Big) {\rm \;mag\;Mpc^{-3}}.
    \label{eq:magdensity}
\end{equation}
Here, $\phi(M)$ is the DPL function of Eq (\ref{eq:dpl}), $M_1=-23$ mag is the bright magnitude limit, and $M_2=-17$ mag is the faint magnitude limit. This value of $M_2$ is chosen so that $\rho_{UV}$ results can be compared easily with other studies which also have the same faint-end limit.

We convert the specific UV luminosity densities to SFRDs using the relationship from \citet{Madau2014},
\begin{equation}
    \dot{m}_{\ast} = \kappa_{UV} \times \rho_{UV},
\end{equation}
where, $\kappa_{UV} = 1.15\times10^{-28} {\rm \; M_{\odot} \; year^{-1} \; erg^{-1} \; s \; Hz}$ is determined from local Universe empirical calibrations for a Salpeter IMF. This provides SFRDs for $4 \leq z \leq 14$.

\begin{figure*}
    \centering
    \includegraphics[width = 0.6\textwidth]{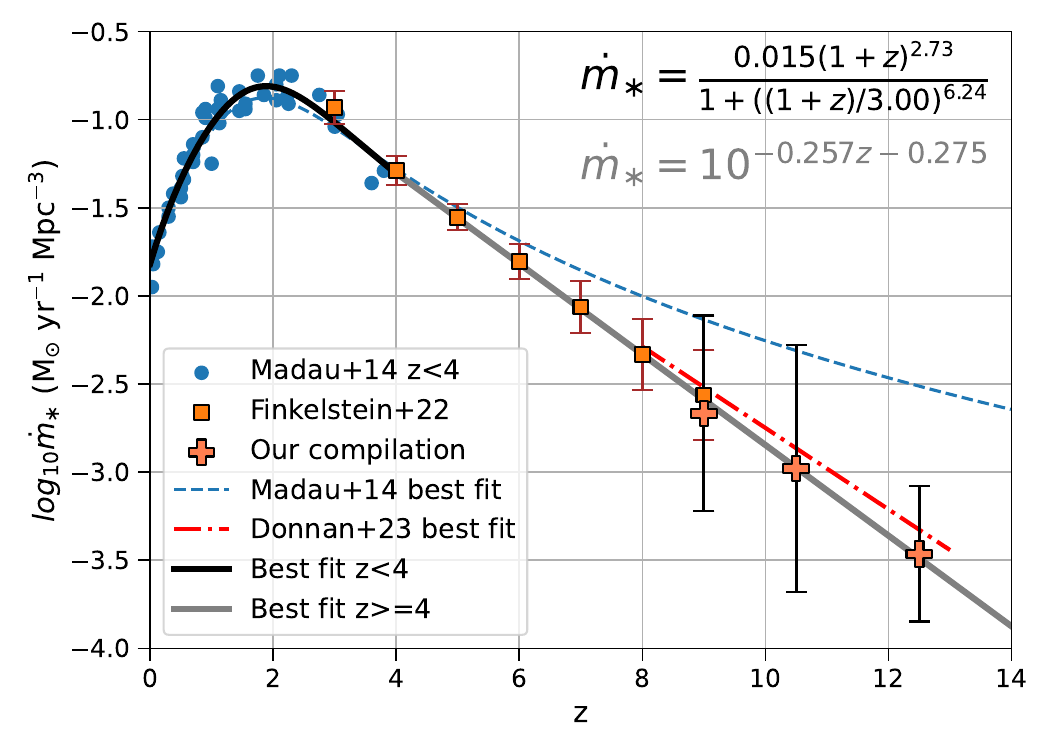}
    \caption{The cosmic star formation rate density upto redshift 14. The high redshift values have been determined using a compilation of results from several JWST studies. The $z>8$ data points agree within the error bars with the evolution determined by \citet{Donnan2023a}. It is clear that high-z star formation rate density may be much lower than estimated by \citet{Madau2014}.}
    \label{fig:sfrd_evol}
\end{figure*}

To get a global evolution of the SFRD across cosmic time, $0 < z < 4$ data from \citet{Madau2014} is fit with the functional form noted in their paper, using maximum likelihood estimation, while $4 \leq z \leq 14$ data are fit with a straight line for log$_{10}\dot{m}_{\ast}$, using nonlinear chi-squared minimization. This cosmic evolution is described as (see Figure \ref{fig:sfrd_evol}),
\begin{equation}
    \dot{m}_{\ast}(z) = 
    \begin{cases}
        0.015 \dfrac{(1+z)^{2.73}}{1 + [(1+z)/3]^{6.24}} & \text{if } 0 < z < 4,\\
        10^{-0.257z - 0.275} & \text{if } 4 \leq z \leq 14.
    \end{cases}
    \label{eq:sfrd}
\end{equation}

\section{CHIME/FRB Catalog 1 data and its baseband data update} \label{sec:cat1_des}

To study the intrinsic properties of the FRB population, \citet{CHIME2021} derive a sub-sample of 265 bursts from the full catalog. Catalog 1 has 536 bursts, wherein multi-component bursts are listed separately, thereby making 600 sub-bursts. The first step is to remove all except the last sub-burst of each multi-component burst. Bursts that are excluded further are sub-optimal or have exposure issues (marked with $\rm excluded\_flag = 1$), are detected in the far-side lobes, have DM $< 1.5\times\rm{max}(\rm{DM_{NE2001}, DM_{YMW16}})$,  $\rm bonsai$ SNR $< 12$, have scattering time longer than 10 ms at 600 MHz, or are bursts subsequent to the first burst detected for a repeater.
% \sout{Those that are excluded include sub-optimal bursts, bursts that have low signal-to-noise ratios (SNRs), high} \sout{scattering times, low DMs, or are bursts detected after the first burst from repeating sources, and} 
The selection process\footnote{Code to extract the 265 burst sample is available at: \url{https://github.com/kaitshin/CHIMEFRB-Cat1-Energy-Dist-Distrs/}} is detailed in \citet{CHIME2021}. 
Some studies (e.g. \citealt{Tang2023}) consider only non-repeating FRBs to constrain the energy distribution. However, following the original sample selection methodology and the reasons outlined in \citet{Shin2023}, we keep the first detected bursts of repeaters surviving the burst selection criteria for analysis.
% , and not technically, FRB events, even though we refer to them as FRB events at several places. 
Individual FRB repeaters may have their own energy distributions, depending on properties intrinsic to the source. This is why pointed observations of repeaters with sensitive telescopes such as FAST are often used to constrain individual source energy distributions (e.g., \citealt{Zhang2022_1, Konijn2024}).

Most studies use the 1D distributions of the observed fluence and DM to match corresponding distributions generated theoretically. However, the fact that fluence is dependent on redshift, which, in turn, is highly correlated with DM, introduces correlations between fluence and DM that cannot be resolved by analyzing 1D distributions (see Figure \ref{fig:scatter_dm_fluence}(a) in Appendix \ref{app:corr_baseband}). Moreover, the determined fluence values in Catalog 1 are highly uncertain and represent lower limits to the actual fluence values \citep{CHIME2021}. This is because the bursts are assumed to be detected along the meridian of the primary beam, and off-meridian attenuation is not accounted for. Using the CHIME injection system \citep{Merryfield2023}, \citet{CHIME2021} also show that the injected fluence values of simulated FRBs and the SNRs recovered by the CHIME/FRB pipeline are correlated. Due to this, SNR has been used as a proxy for fluence in \citet{Shin2023}, who model the joint distribution of DM and SNR to determine the posterior distributions of energy and redshift distribution parameters (among other things). However, there is a very significant scatter in the correlation of injected fluence and SNR, which means that even SNR is not a completely reliable proxy for fluence (discussed more in Section \ref{sec:Edist_and_rate}).

To obtain reliable fluences, we employ the baseband data update, which has information for 140 bursts forming a subset of Catalog 1 \citep{BasebandCatalog1_2023}. The entire baseband catalog is cross-referenced with the set of 265 FRBs extracted from Catalog 1, which yields 93 bursts. There is one repeater in the list of 265 FRBs, whose first burst detected in Catalog 1 is not its first burst detected in Baseband Catalog 1. Including its first burst detected in Baseband Catalog 1, makes our final sample size to be 94 bursts.\footnote{Code to extract the sample of 94 FRBs is available at: \url{https://github.com/omguptaup/frb-cosmic-evol}} 
% \sout{With the selection criteria used to sub-sample Catalog 1, but taking the first occurring burst from each repeater that} \sout{is present in the Baseband catalog, the number of remaining usable bursts is 94.} 
The updated fluences in the baseband sample are accurate to within 10\% \citep{BasebandCatalog1_2023}, but the number of bursts that can be analyzed decreases to about a third of the initial CHIME subsample of 265 bursts.

\section{The theoretical model} \label{sec:theory_model}

\subsection{The FRB Energy and Redshift Distribution}
We choose to work with FRB specific energies\footnote{Specific energy or spectral energy refers to the energy per unit frequency $dE/d\nu$, denoted as $E_{\nu}$.} following the methodology of \citet{Beniamini2021}, and assume that the distribution of FRB number fraction with isotropic-equivalent spectral energy at a frequency $\nu$ (referred hereinafter as the energy distribution) is defined by a Schechter function \citep{Lu2019, Lu2020, Shin2023}. The probability density function for isotropic FRB specific energy $E_{\nu}$ in the comoving frame of the burst is
\begin{equation}
    P(E_{\nu}) dE_{\nu} = \frac{1}{\Gamma(\gamma + 1, E_{\nu}^{\rm pivot}/E_{\rm char, \nu}) E_{\rm char,\nu}} \Big( \frac{E_{\nu}}{E_{\rm char,\nu}} \Big)^{\gamma} {\rm exp} \Big( \frac{-E_{\nu}}{E_{\rm char,\nu}} \Big) dE_{\nu}, \;\;\;{\rm for}\; E_{\nu} \geq E_{\nu}^{\rm pivot}.
    \label{eq:frbEdist_schechter}
\end{equation}
In this equation, $E_{\rm char, \nu}$ is the characteristic specific energy where the exponential cutoff begins. Eq. (\ref{eq:frbEdist_schechter}) needs to be calibrated for a single frequency, which for CHIME data is $\nu = 600$ MHz in the comoving frame in each redshift bin. The calibrated energy distribution is assumed to be valid at all frequencies within a reasonable range. The energy distribution in Eq. (\ref{eq:frbEdist_schechter}) is normalized 
to 1, and the normalization factor is defined as, 
\begin{equation}
    \Gamma \Big(\gamma+1, \frac{E_{\nu}^{\rm pivot}}{E_{\rm char, \nu}} \Big) = \int_{E_{\nu}^{\rm pivot}}^{\infty} \frac{1}{E_{\rm char,\nu}} \Big( \frac{E_{\nu}}{E_{\rm char,\nu}} \Big)^{\gamma} {\rm exp} \Big( \frac{-E_{\nu}}{E_{\rm char,\nu}} \Big) dE_{\nu}
    \label{eq:Gamma_func}
\end{equation}
In both Eq. (\ref{eq:frbEdist_schechter}) and Eq. (\ref{eq:Gamma_func}), the pivot energy $E_{\nu}^{\rm pivot}$ doubles the role as a convenient value for quoting results and also as a marker of our confidence in how low the assumed form of the energy distribution extends. For its former use case, the choice of the pivot energy $E_{\nu}^{\rm pivot}$ does not affect results. 
For the latter use-case, the motivation for its value is extensively discussed in Section \ref{sec:discussion_Epivot}, and we choose $E_{\nu}^{\rm pivot} = 10^{30}$ \energyunit. 

We assume that the spectral energy number distribution does not evolve with redshift. The FRB rate at a certain energy per unit comoving volume at redshift $z$, then, is
\begin{equation}
    \frac{dN(z, E_{\nu})}{dt_{\rm com} dV}  = \Phi_0 P(E_{\nu}) \psi_{\ast}(z) dE_{\nu},
    \label{eq:com_frb_rate}
\end{equation}
where, $\Phi_0$ is the volumetric rate of FRBs at $z=0$ and at rest-frame frequency of 600 MHz in Mpc$^{-3}$ year$^{-1}$, and, $\psi_{\ast}$ is a redshift-evolving property of the population of stars that are believed to form the massive compact objects that produce FRBs. In Section \ref{sec:model_description}, we will try two redshift-dependent functions for $\psi_{\ast}$. No evolution of the IMF with redshift is assumed, even though it is well-understood that the IMF will become top-heavy with increasing redshift (e.g., \citealt{Finkelstein2023, Bromm2004}).

As CHIME observations are made at $\nu_0$, the corresponding comoving frequency is $\nu_{\rm com} = (1+z) \nu_0$ at a redshift of $z$\footnote{There are 3 frequencies identified in this work: $\nu$, $\nu_0$ and $\nu_{\rm com}$. $\nu_0$ is the observing frequency (= 600 MHz for CHIME), and appears only as a subscript on fluence, which is an observational quantity. $\nu$ and $\nu_{\rm com}$ are frequencies in the comoving frame. The value of $\nu=\nu_0=600$ MHz during calibration, such that an energy distribution at a single frequency can be constrained. Whereas, we set $\nu=\nu_{\rm com}$ post-calibration, assuming the constrained energy distribution to be valid at all comoving frequencies, with this usage beginning in Section \ref{sec:results}.}. For a given redshift $z$, our aim is to quantify the rate of FRBs above the detector's fluence threshold $F_{\nu_0}^{\rm th}$. The isotropic specific energy released at $\nu_{\rm com}$ is $E_{\nu_{\rm com}} = F_{\nu_0} \times 4\pi d_{\rm com}^2$, where $d_{\rm com}$ is the comoving radial distance to $z$.
Eq (\ref{eq:frbEdist_schechter}) is not applicable to $E_{\nu_{\rm com}}$  because the energy distribution is being calibrated at some specific $\nu$ in the comoving frame.
Consequently, we must assume either that there exists an intrinsic statistical spectral energy distribution (SED; distribution of energy with frequency) for FRBs, or that the energy distribution parameters evolve with frequency. Choosing the latter option makes the problem more complex as $\gamma(\nu)$ and $E_{\rm char, \nu}(\nu)$ will introduce more than 2 parameters that would need to be constrained. We choose the former option, taking a power-law SED that gives a relation between the specific energies of bursts happening at different frequencies as,
\begin{equation}
    E_{\nu} = E_{\nu_{\rm com}} \Big( \frac{\nu}{\nu_{\rm com}} \Big)^{\alpha}.
    \label{eq:energy_pl}
\end{equation}
With this, we end up having three free parameters, namely $\gamma$, $E_{\rm char, \nu}$, and $\alpha$.
The power-law statistical SED is the spectral slope obtained for an ensemble of FRBs in the comoving frame across a wide range of frequencies. The individual burst spectrum may be different than the statistical one. The choice of a statistical SED is motivated by the observations of FRBs from 8 GHz down to 110 MHz \citep{Gajjar2018, Pleunis2021b} and the fact that even repeaters, which generally exhibit narrow-band bursts \citep{Curtin2024}, are detected across a wide range of frequencies using different telescopes. The energy distribution may be evolving with rest-frame frequency/redshift (source evolution), but this scenario cannot be probed presently.
Since, we observe fluence and not specific energy, we can relate the two using,
\begin{equation}
    E_{\nu} = \frac{4\pi d_{\rm com}^2(z)}{(1+z)^{\alpha}} F_{\nu_0}.
    \label{eq:Eflu_rel}
\end{equation}
From this and Eq (\ref{eq:frbEdist_schechter}), we can deduce that $P(E_{\nu})dE_{\nu} = P(F_{\nu_0})dF_{\nu_0}$. Assuming that Eq (\ref{eq:com_frb_rate}) holds at all redshifts, the differential FRB rate per unit comoving volume at a given fluence in the observer frame,
\begin{equation}
    \frac{dN(z,F_{\nu_0})}{dt dV} = \frac{dN(z, F_{\nu_0})}{dt_{\rm com}dV} \frac{dt_{\rm com}}{dt} = \frac{dN(z, F_{\nu_0})}{dt_{\rm com} dV} \frac{1}{1+z} = \Phi_0 P(E_{\nu}) \psi_{\ast}(z) \frac{4\pi d_{\rm com}^2(z)}{(1+z)^{1+\alpha}} dF_{\nu_0}.
\end{equation}
Here, the all-sky differential comoving volume element $dV = 4\pi d_{\rm com}^2(z)c/H(z) dz$, and $H(z)$ is the Hubble constant at redshift $z$. With this, we define the FRB rate function as,
\begin{equation}
    R(z, F_{\nu_0}) = \frac{dN(z,F_{\nu_0})}{dt dz dF_{\nu_0}} = \Phi_0 P(E_{\nu}) \psi_{\ast}(z) \frac{16\pi^2 d_{\rm com}^4(z)}{(1+z)^{1+\alpha}} \frac{c}{H(z)}.
    \label{eq:diffrate}
\end{equation}
We assume a flat $\Lambda$-CDM cosmology with Planck-2018 cosmological parameters \citep{Planck2020}. It is important to note that Eq. (\ref{eq:diffrate}) is defined at $\nu=600$ MHz in the comoving frame while calibrating the energy distribution.

\subsection{Generating the Fast Radio Burst Dispersion Measure Distribution} \label{sec:dmsection}

The distance indicator for FRBs is the Dispersion Measure (DM). But the observed DM is composed of electron column density components along the line-of-sight belonging to the interstellar medium of the Milky Way (MW-ISM), the Milky Way halo, the intergalactic medium (IGM) between the host galaxy and the Milky Way with the contribution of intervening galaxy halos, the host galaxy halo and ISM, and finally, the immediate environment of the FRB source. The observed DM then can be written as a sum of these components,
\begin{equation}
    \rm DM = DM_{\rm MW, ISM} + DM_{\rm MW, halo} + DM_{\rm IGM} + DM_{\rm host},
    \label{eq:dm_components}
\end{equation}
where, the observed and comoving values of $\rm DM_{host}$ are related by,
\begin{equation}
    {\rm DM_{host}} = \frac{\rm DM_{\rm host, com}}{1+z} = \frac{\rm DM_{\rm host, ISM, com} + DM_{\rm host, halo, com}}{1+z}.
    \label{eq:dmhost}
\end{equation}
Assuming that DM associated with the immediate environment of the source is typically small enough that it can be folded into the uncertainties of the other host components, we neglect DM$_{\rm source}$. 
Furthermore, each component of Eq. (\ref{eq:dm_components}) has its own distribution of values, which may depend on redshift. 

DMs contributed by MW-ISM are well-modelled along different sightlines using the pulsar-calibrated Galactic electron density models NE2001 \citep{Cordes2002ne2001} and YMW16 \citep{Yao2017}, the latter of which we have not used here. 
NE2001 has large uncertainties, but is regularly used to estimate Milky Way DMs for FRBs. The MW-ISM DM estimates determined using NE2001 for all bursts in our sample, have been provided by \cite{CHIME2021}. We fit a lognormal distribution on the ${\rm DM_{MW, ISM}}$ estimates of all 265 Catalog 1 bursts, which serves as its crude probability density function. In no way does it represent the actual distribution of ${\rm DM_{MW, ISM}}$, but such a distribution is constructed because the calibration procedure requires sampling from an observed distribution.  
The DM contributed by the Milky Way halo is assumed to be 50 pc cm$^{-3}$, which is consistent with upper limits between 52-111 pc cm$^{-3}$ derived using various models of the ionized Milky Way halo \citep{Cook2023}.

\subsubsection{Estimating Intergalactic Medium Dispersion Measure Contributions} \label{sec:dmigm}

Most of the baryons in the Universe are believed to be in a highly diffuse state residing in the IGM, and forming roughly 83\% of the total baryon budget at $z\sim 0$ \citep{Fukugita1998, Cen2006, Nicastro2018, Macquart2020}. For a long time, a large fraction of these IGM baryons were too hard to detect \citep{McQuinn2016}, but FRBs have provided a promising avenue to locate these ``missing" baryons \citep{Macquart2020, Yang2022}. This is primarily because at $z< 3$, the IGM baryons are almost entirely in an ionized phase, causing the contribution of the IGM to the total DM to be an increasing function with FRB distance. For cosmic times earlier than the second helium reionization, the ionized fraction of baryons in the IGM decreases. The fraction of ionized baryons declines steadily to zero during the epoch of hydrogen reionization and the first helium reionization, but that period is beyond the observable regime of CHIME, and hence will be discussed later.

The average ${\rm DM}_{\rm IGM}$ can be fairly well-modelled using the following,
\begin{equation}
    \langle {\rm DM_{IGM}} \rangle = c\int_0^z dz' \frac{\langle n_e(z') \rangle}{(1+z')^2H(z')}
    \label{eq:dmigm_temp}
\end{equation}
For the flat $\Lambda$-CDM cosmology we have adopted, with matter and dark energy as its dominant components, the Hubble constant,
\begin{equation}
    H(z) = H_0 \displaystyle\sqrt{\Omega_{m0}(1+z)^3 + \Omega_{\Lambda0}},
    \label{eq:hubble_constant_z}
\end{equation}
where, $\Omega_{m0}$ and $\Omega_{\Lambda 0}$ are the dimensionless matter and dark energy densities respectively, and $H_0$ is the present value of the Hubble constant. The average electron density in a comoving volume element is given by 
\begin{equation}
    \langle n_e(z) \rangle = f_d(z) \rho_b(z) \Big[ \frac{Y_{\rm H}}{m_p}f_{\rm HII}(z) + \frac{Y_{\rm He}}{4m_p}\xi_{\rm He}(z)\Big],
\end{equation}
where $f_d$ is the fraction of baryons in the diffuse ionized IGM, $\rho_b(z) = 3H_0^2\Omega_{b0} (1+z)^3/8\pi G$ is the comoving baryon density, $m_p$ is the mass of proton, $Y_{\rm H}\approx 0.75$ and $Y_{\rm He}\approx 0.25$ are the mass fractions of hydrogen and helium out of all cosmic baryons, respectively, $f_{\rm HII}$ is the fraction of ionized hydrogen in the IGM, and $\xi_{\rm He}$ is the average number of electrons contributed by helium per He nucleus. Thus, Eq (\ref{eq:dmigm_temp}) can be written as,
\begin{equation}
    \langle {\rm DM_{ IGM}} \rangle = \frac{ 3 c H_0 \Omega_{b0}}{8\pi G m_p} \int_0^z dz' \frac{f_d(z')(1+z')}{[\Omega_{m0}(1+z')^3 + \Omega_{\Lambda0}]^{1/2}} \Big[ Y_{\rm H}f_{\rm HII}(z') + \frac{Y_{\rm He}}{4}\xi_{\rm He}(z')\Big].
    \label{eq:dmigm}
\end{equation}

We model the average DM$_{\rm IGM}$ by adapting the publicly available FRB repository\footnote{\url{https://github.com/FRBs/FRB}} encoded in Python \citep{Macquart2020}. In this formulation, $f_d$ is calculated by subtracting from one the fraction of baryons locked up in stars, in stellar remnants and as neutral gas within galaxies. The redshift evolution of stars comes from calculating the stellar mass density from the star formation rate in Eq (\ref{eq:sfrd}), and is detailed in Section \ref{sec:model_description}. Stellar remnants, such as white dwarfs, neutron stars and black holes, are considered to be 30\% of the stellar mass \citep{Fukugita2004}, without any cosmic evolution of this fraction. For the ISM component, \citet{Macquart2020} adopt the estimate of $m_{\rm ISM}/m_{\ast}=0.38$ \citep{Fukugita2004} at $z=0$, and interpolate it quadratically to $m_{\rm ISM}/m_{\ast}=1.0$ at $z=1$. This ratio is assumed to remain constant for $z>1$.

Different FRB sightlines would probe different structures in the IGM formed by the collapse of matter into large-scale structures, such as cosmic sheets, filaments and voids, in addition to foreground intervening halos. As such, there is a distribution of DM$_{\rm IGM}$ values at any given redshift, with the mean estimated by Eq (\ref{eq:dmigm}). To estimate the scatter in DM$_{\rm IGM}$, we use the distributions found by \citet{Jaroszynski2019} at different redshifts, using Illustris-3 cosmological simulations. It is noted that these distributions are skewed with the tail towards larger DMs, and the DM$_{\rm IGM}$ distribution becomes more symmetric at increasing redshifts. We use a four-parameter family of distributions described in \citet{Jones2009} to generate skewed distributions whose first four moments match those found by \citet{Jaroszynski2019}. The parameters are described by fitting functions that can generate a DM$_{\rm IGM}$ distribution at any redshift, and the process for finding them is discussed in Appendix \ref{sec:appb}. 

The scatter in the DM$_{\rm IGM}$ distribution closely follows the relation $\sigma_{\rm DM, IGM} = 0.13 {\rm DM}_{\rm IGM} / (1+z)$ \citep{Beniamini2021}. This result agrees with the lower-end of the range of values predicted by \citet{McQuinn2014} at $z \simeq 0.5-1$, but is smaller by more than $\sim 35\%$ from the values predicted by \citet{Jaroszynski2020}. The evolution of $\langle {\rm DM_{IGM}} \rangle$ with redshift in \citet{Jaroszynski2020} conforms very closely to the evolution predicted with $f_d \approx 1$. This means that at least some part of the excess IGM inhomogeneity in that work is due to an inflated fraction of baryons in the IGM. The distributions found in a more recent work using IllustrisTNG \citep{Zhang2021b} are also wider than the ones we use, and how the results are affected by using constraints from IllustrisTNG is discussed in Section \ref{sec:discussion_robust}.

\subsubsection{Estimating Dispersion Measure Contributions from the Host Galaxy} \label{sec:dmhost}
Finally, the host galaxy and its halo add their own contribution to the total DM. The column density of electrons along the line-of-sight is mostly dependent on the location of the FRB in its host galaxy, and thus, on the path the FRB takes along the warm ionized interstellar medium of the host, and then, through the halo. The amount or density of gas in the ISM also affects DM$_{\rm host}$. Consequently, a large scatter in the amount of host contributions is observed. Total DMs that are well in excess of $\langle {\rm DM_{IGM}}\rangle$ are exhibited by the repeating FRBs, FRB 20121102A \citep{Tendulkar2017} and FRB 20190520B \citep{Niu2022}, which point to large values of DM$_{\rm host, ISM+halo}$ + DM$_{\rm source}$.

Clarifying the notational use of $P(\rm Quantity)$ throughout this work to denote differential probability with respect to the corresponding quantity, the host's total contribution in its comoving frame is generally modelled as a lognormal distribution (e.g., \citealt{Macquart2020, James2022b, Shin2023}),
\begin{equation}
    P({\rm DM_{host, com}|z}) = \frac{1}{\rm DM_{host, com}} \frac{1}{\sigma_{\rm host}(z) \sqrt{2\pi}} {\rm exp} \Big[ -\frac{(ln {\rm DM_{host, com}} - \mu_{\rm host}(z))^2}{2\sigma_{\rm host}^2(z)} \Big],
\end{equation}
that is parameterized by $\mu_{\rm host}$ and $\sigma_{\rm host}$. These parameters are related to the median as $\rm exp(\mu_{\rm host})$ and to the standard deviation of the distribution as $[{\rm exp}(2\mu_{\rm host} + \sigma_{\rm host}^2) (e^{\displaystyle \sigma_{\rm host}^2} - 1)]^{1/2}$. Unlike previous papers, which assume a constant median and standard deviation, we consider the evolution of the ionized gas content of galaxies over cosmic time.

Using Illustris-TNG100-1 cosmological simulations, \citet{Mo2023} find the evolution of the median and $\sigma_{\rm host}$ of the DM$_{\rm host}$ distribution for $z \leq 2$ \footnote{This data, underlying Figure 10 of \citet{Mo2023}, is kindly provided by Weishan Zhu.}. They estimate this evolution for two kinds of FRB populations, one tracking the star-formation rate and the other tracking the stellar mass density. This allows us to use a separate evolution track for host galaxies corresponding to the function being implemented as $\psi_{\ast}$ in Eq (\ref{eq:com_frb_rate}). We find that the median values are fit well with a hyperbolic-tangent function, which has the advantage over their power-law fit that it does not keep increasing with redshift. At $z \gtrsim 3$, when the decreasing size of galaxies dominates over the gas mass, $\rm DM_{host, com}$ should start decreasing. Combined with the effect of observed contribution being smaller by a factor of $1+z$, the error introduced by assuming a constant $\rm DM_{host, com}$ at high redshift is not significant. Moreover, DMs are used just for calibration, so how their distribution evolves at redshifts higher than those in the observed sample is inconsequential. The fit for the median is given as,
\begin{equation}
    {\rm exp(\mu_{ host})} = 
    \begin{cases}
        155.9 + 179.8 \times {\rm tanh}(1.02z) & \text{when SFR-tracking, and}\\
        59.6 + 293.0 \times {\rm tanh}(0.61z) & \text{when SMD-tracking.}\\
    \end{cases}
\end{equation}
However, $\sigma_{\rm host}$ is obtained by linear interpolation for both cases, and is kept constant for $z\geq 2$. This leads to
\begin{equation}
    P(\rm DM_{host}|z) = P({\rm DM_{host, com}|z}) \frac{d\rm DM_{host,com}}{d\rm DM_{host}} = P({\rm DM_{host, com}|z}) \times (1+z)
    \label{eq:P_dmhost}
\end{equation}

\subsection{Models Describing Redshift Evolution} \label{sec:model_description}
Based on the discussion in Section \ref{sec:intro}, FRB redshift evolution is yet to be determined. Still, there are good reasons reliant upon observations and other phenomena, like Gamma-ray bursts, why it could possibly be well-approximated by one of a few known models or functional forms.
If FRBs are primarily produced by sources generated in core-collapse supernovae, then $\psi_{\ast}(z)$ would very closely track the star-formation rate $\dot{m}_{\ast}(z)$ empirically determined in Eq (\ref{eq:sfrd}). If the progenitors are part of older systems with significant delays from star formation, then $\psi_{\ast}(z)$ would follow the stellar mass density, defined as \citep{Madau2014},
\begin{equation}
    m_{\ast}(z) = (1-R) \int_z^{\infty} \frac{\dot{m}_{\ast}(z')}{(1+z')H(z')} dz',
\end{equation}
where, $H(z)$ is defined in Eq (\ref{eq:hubble_constant_z}), and $R=0.27$ is the “return” fraction for Salpeter IMF. $R$ is the mass fraction of massive stars of each generation that is returned to the ISM and IGM after their death. If both young and old systems contribute to the total FRB source population, then $\psi_{\ast}$ can be defined as a linear combination of the two, 
\begin{equation}
    \psi_{\ast}(z, f_{\rm Y}) = f_{\rm Y} \frac{\dot{m}_{\ast}(z)}{\dot{m}_{\ast}(z=0)} + (1-f_{\rm Y})\frac{m_{\ast}(z)}{m_{\ast}(z=0)}.
    \label{eq:sfrsmd_hybrid}
\end{equation}
Here, $f_{\rm Y}$ is the fraction of FRB sources belonging to the young population, and serves as another free parameter for models of the SFR-SMD hybrid kind.

In addition to the SFR-SMD hybrid model, we also try a model with a constant delay time of $\tau$ Gyr with respect to SFR. This model is used to estimate the average delay time of the formation of FRB sources from star formation. Like in \citet{Shin2023}, it is also used to assess the credibility of models with delay-time distributions, such as the ones used in \citet{Zhang2022} borrowed from Gamma-ray burst literature. We also test two delay time distributions that do not introduce additional parameters to constrain, in Appendix \ref{sec:delay_time}. For the constant delay time case,
\begin{equation}
    \psi_{\ast}(z, \tau) = \frac{\dot{m}_{\ast}(Z_a(t_{\rm lb}(z) + \tau))}{ \dot{m}_{\ast}(z=0)},
    \label{eq:constdelaytime}
\end{equation}
where, $t_{\rm lb}(z)$ is the lookback time at redshift $z$, and $Z_a$ is a function which converts a given lookback time to redshift. $\tau$ is varied from 0 to 12 Gyr. 

\subsection{Determining the Joint Distribution of Dispersion Measure and Fluence} \label{sec:dist_dmnF}
Our goal has been to find the theoretical probability distribution function of two FRB observables, namely the DM and the fluence. It can be calculated as follows,
\begin{equation}
    P({\rm DM}, F_{\nu_0}) = \int_{z_{\rm min}}^{z_{\rm max}} dz P({\rm DM}|z) \times R(z, F_{\nu_0}),
    \label{eq:P_dmnF}
\end{equation}
and, whose units are all-sky year$^{-1}$ DM$^{-1}$ $F_{\nu_0}^{-1}$. The integral is evaluated on a grid of $z=0-8.00$, linearly spaced with a step size of 0.01. $R(z, F_{\nu_0})$ assumes the value at the center of the bin, using Eq (\ref{eq:diffrate}), and depends on the chosen redshift evolution model $\psi_{\ast}$. The full redshift range $z\in [0,14]$ is not chosen, as mentioned above, because the probabilities for our observed sample are expected to become negligible, and thus, computing Eq (\ref{eq:P_dmnF}) over a shorter redshift range provides a decreased computational requirement.

Since, DM is comprised of several components, some of which have their own distributions, convolution is used to determine $P({\rm DM}|z)$. It can be computed as,
\begin{eqnarray}
    &P({\rm DM}|z) = P({\rm DM_{EG} = DM-DM_{NE2001} - DM_{MW, halo}}|z) \nonumber \\
    &= \int d{\rm DM_{IGM}} P({\rm DM_{host} = DM-DM_{NE2001} - DM_{MW, halo} - DM_{IGM}}|z) P({\rm DM_{IGM}|z}).
    \label{eq:P_DMgz}
\end{eqnarray}
As discussed in Section \ref{sec:dmsection}, ${\rm DM_{MW,halo}}=50$ pc cm$^{-3}$, and $\rm DM_{NE2001}$ is the NE2001 \citep{Cordes2002ne2001} estimate of the Milky Way ISM electron column density. Eq (\ref{eq:P_DMgz}) will be used to calculate Eq (\ref{eq:P_dmnF}) for each FRB in our sample. However, the log-likelihood function defined in Section \ref{sec:posterior_likelihood} requires the computation of Eq (\ref{eq:P_dmnF}) in DM and $F_{\nu_0}$ bins. In a bin, which can have multiple FRBs, $\rm DM_{MW, ISM}$ is sampled from the distribution of possible values given by a lognormal fitted to the NE2001 estimates of FRBs in the selection-correction sub-sample of Catalog 1.
\begin{eqnarray}
    P({\rm DM_j}|z) &= \int d{\rm DM}_{\rm MW,ISM} P({\rm DM_{MW,ISM}}) P({\rm DM_{EG} = DM_j-DM_{MW, ISM} - DM_{MW, halo}}|z) \nonumber \\
    &= \int d{\rm DM}_{\rm MW,ISM} P({\rm DM_{MW,ISM}}) \int d{\rm DM_{IGM}} P({\rm DM_{host} = DM_j-DM_{MW, ISM}} \nonumber \\ & {\rm - DM_{MW, halo} - DM_{IGM}}|z) P({\rm DM_{IGM}|z}),
    \label{eq:P_DMgz_bin}
\end{eqnarray}
where, $\rm DM_j$ represents the value in the middle of the j-th DM bin.
For the SFR-SMD hybrid model, Eq (\ref{eq:P_dmhost}) needs to be modified to account for the fraction $f_{\rm Y}$ being SFR-tracking hosts and $(1-f_{\rm Y})$ being SMD-tracking hosts. Consequently,
\begin{equation}
    P({\rm DM_{host}}|z) = [f_{\rm Y}\times P({\rm DM_{host,com}}|z,{\rm SFR}) + (1-f_{\rm Y})\times P({\rm DM_{host,com}}|z,{\rm SMD})] \times (1+z).
\end{equation}
For the constant delay time model, Eq (\ref{eq:P_dmhost}) is used with just SMD tracking parameters.

\subsection{Instrumental Effects and CHIME's Observation Function}

Being a real telescope, CHIME suffers from instrumental biases and selection effects that need to be carefully modelled, to be able to calibrate an FRB event rate model with the data. The FRB injection pipeline \citep{Merryfield2023} is used to account for selection effects, that are most strongly reflected in a bias against FRBs with low fluences or with high-scattering times \citep{CHIME2021}. 
\citet{CHIME2021} outlines the methodology used to calculate the observation function\footnote{The observation function is kindly provided by Kiyoshi Masui.} $P({\rm SNR | DM}, F_{\nu_0})$, defined as the probability of observing a particular SNR for an injected burst, given the burst's DM and fluence $F_{\nu_0}$.

According to the CHIME pipeline, since FRBs are said to be confidently detected for SNR $\geq 12$, the conditional probability of detection is $P({\rm SNR}\geq 12 | {\rm DM}, F_{\nu_0}) = \int_{12}^{\infty} P({\rm SNR | DM}, F_{\nu_0}) d{\rm SNR}$. The observation function is also defined above this SNR limit, and is the reason why not all bursts in the baseband catalog can be used.
Because the injected events dictate the value of $P({\rm SNR \geq 12 | DM}, F_{\nu_0})$ in a particular DM and $F_{\nu_0}$ bin, their limited number leads to holes and poorly sampled points in the matrix, especially towards the edges of the sampled parameter space. So, we first use nearest value interpolation on the same grid (\textit{scipy.interpolate.griddata}). Since, the grid on which the observation function is defined is irregular,
we then smooth the interpolated observation function 
using \textit{scipy.interpolate.SmoothBivariateSpline}.
Now, the DM$-F_{\nu_0}$ probability distribution of detectable FRBs can be calculated as follows,
\begin{equation}
    P({\rm SNR \geq 12, DM}, F_{\nu_0}) = P({\rm SNR \geq 12| DM}, F_{\nu_0}) \times P({\rm DM}, F_{\nu_0}).
    \label{eq:joint_dist}
\end{equation}
\begin{equation}
    P({\rm DM}, F_{\nu_0} | {\rm SNR \geq 12}) = \frac{P({\rm SNR \geq 12, DM}, F_{\nu_0})}{P({\rm SNR \geq 12})} = \frac{P({\rm SNR \geq 12, DM}, F_{\nu_0})}{\sum_{\rm j} \sum_{\rm k} P({\rm SNR \geq 12, DM_j}, F_{\nu_0, \rm k}) \Delta {\rm DM_j} \Delta F_{\nu_0, \rm k}}.
    \label{eq:condprob}
\end{equation}

\subsection{Obtaining Model Parameter Posteriors with Markov Chain Monte Carlo} \label{sec:posterior_likelihood}
\begin{figure*}
    \gridline{\fig{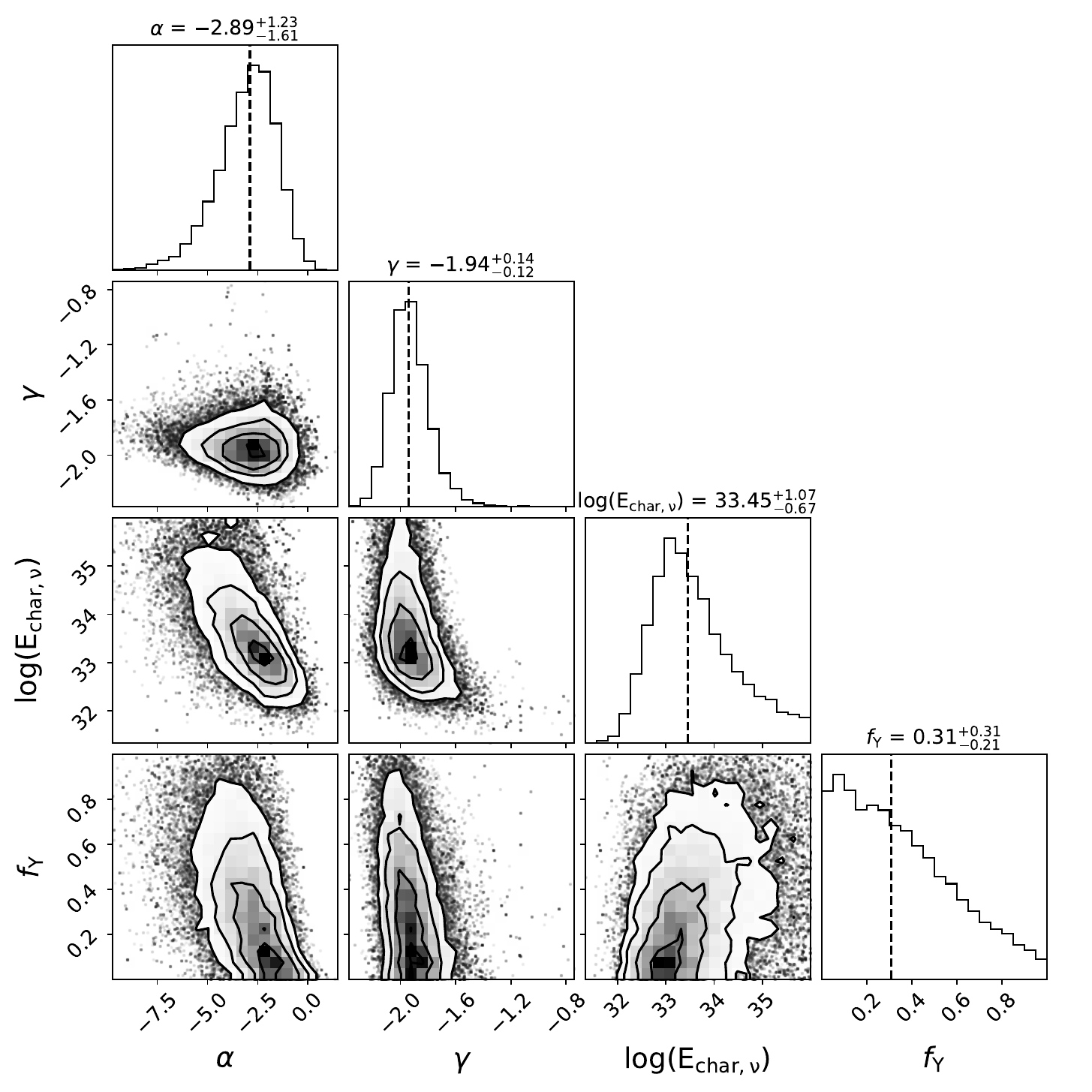}{0.48\textwidth}{(a)}
    \fig{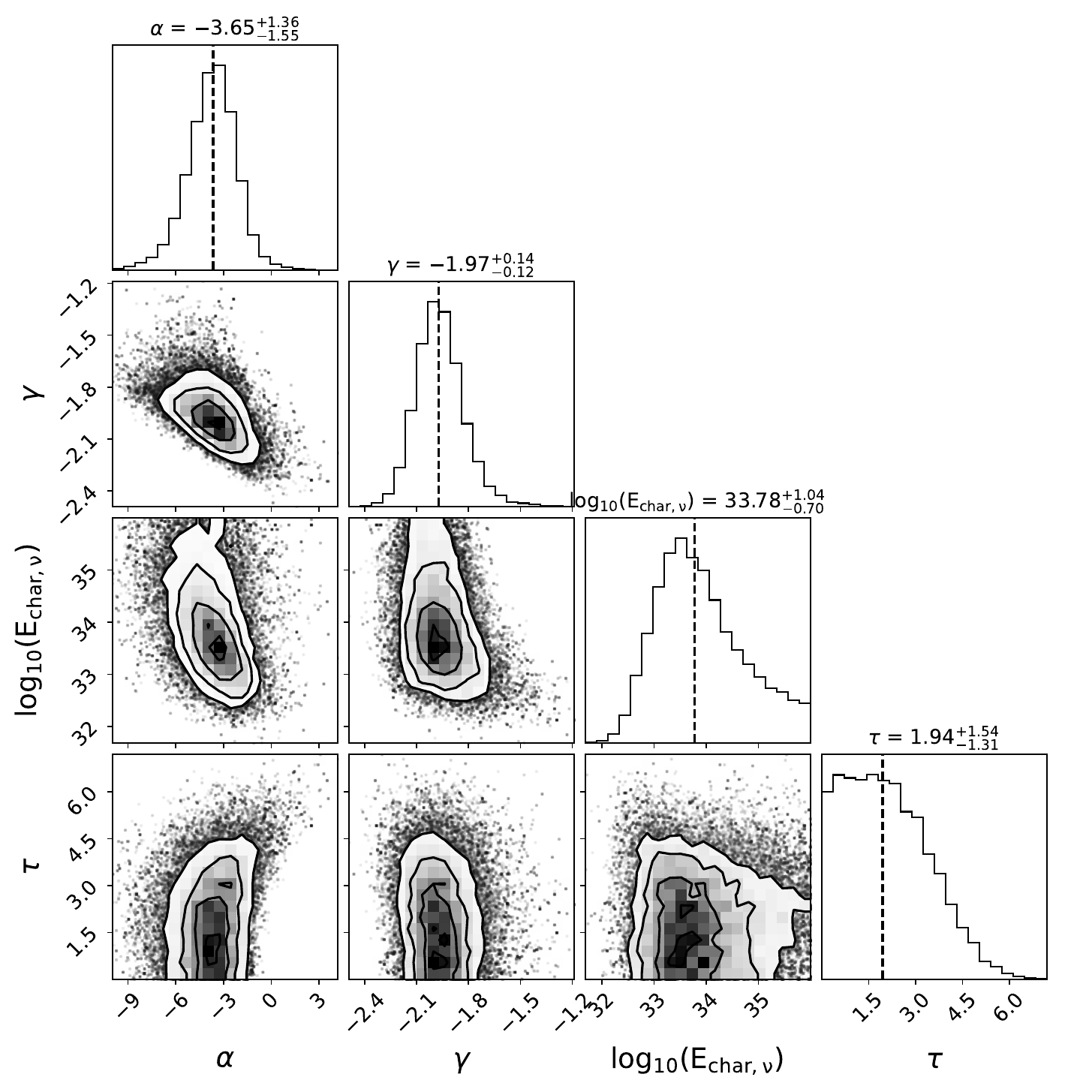}{0.48\textwidth}{(b)}
    }
    \caption{Posterior distributions of (a) the SFR-SMD hybrid model parameters, and (b) the constant delay time model parameters, for the unbinned likelihood defined in Eq (\ref{eq:agEf_likelihood}). 
    }
    \label{fig:cornerplot_param_agEf}
\end{figure*}

Model parameters need to be estimated using the available CHIME baseband data, for which we implement MCMC. Let $i \in [1,94]$, such that ${\rm DM}_i$ and $F_{\nu_0, i}$ denote the $i^{\rm th}$ observed FRB's corresponding DM and fluence, respectively. Because the number of samples is less, it is easy to lose information through binning the data. This is why an unbinned log-likelihood function is used \citep{Zyla2020},
\begin{equation}
    {\rm log}\mathcal{L} = \sum_{i=1}^{94} {\rm log}\, P({\rm DM}_i, F_{\nu_0,i} | {\rm SNR \geq 12}) 
    % = \sum_{i=1}^{94} {\rm log}\,\frac{P({\rm SNR \geq 12, DM}_i, F_{\nu_0,i})}{\sum_{\rm j} \sum_{\rm k} P({\rm SNR \geq 12, DM_j}, F_{\nu_0, \rm k}) \Delta {\rm DM_j} \Delta F_{\nu_0, \rm k}}
    \label{eq:agEf_likelihood}
\end{equation}
Even for a log-likelihood function using observed FRB properties instead of binned data, the calculation of $P({\rm SNR \geq 12})$ requires marginalization of ${\rm DM}$ and $F_{\nu_0}$ in Eq (\ref{eq:joint_dist}). To keep this marginalization economical, the computation of Eq. (\ref{eq:condprob}) is performed on a grid defined over ${\rm DM}=100-5000$ pc cm$^{-3}$ divided into 26 bins and $F_{\nu_0}=1-1500$ Jy ms divided into 40 bins, both spaced logarithmically.
For the SFR-SMD hybrid model, there are four free parameters: $\alpha$, $\gamma$, $E_{\rm char, \nu}$ and $f_{\rm Y}$, all of which are given uniform priors. Similarly, the constant delay time parameters, $\alpha$, $\gamma$, $E_{\rm char, \nu}$ and $\tau$, are also given uniform priors. The MCMC chain is run with 32 walkers for 3000 steps, and the first 100 steps are discarded. Having a chain which is longer than 75 times the integrated autocorrelation length of each parameter, and obtaining an acceptance fraction of $\approx 0.24$ increases our confidence in the convergence of the MCMC chain. The value of the acceptance fraction is optimized by using a combination of the \textit{DEMove} and \textit{DESnookerMove}, instead of the default proposal for updating MCMC walker coordinates \citep{Foreman-Mackey2013}. The posterior distributions of parameters are shown in Figure \ref{fig:cornerplot_param_agEf}, and their median values define our fiducial models. The range of priors and obtained parameter values are listed in Table \ref{tab:dist_params} for both SFR-SMD hybrid as well as constant delay time models.

\begin{deluxetable*}{ccccc} \label{tab:dist_params}
\tablecaption{Redshift and energy distribution parameter constraints.}
\tablehead{\colhead{Parameter} & \multicolumn{2}{c}{SFR-SMD Hybrid Model} & \multicolumn{2}{c}{Constant Delay Time Model} \\ & \colhead{Prior} & \colhead{Posterior} & \colhead{Prior} & \colhead{Posterior} }
\startdata
$\alpha$ & [-10, +5] & $-2.89^{+1.23}_{-1.61}$ & [-10, +5] & $-3.65^{+1.36}_{-1.55}$ \\
$\gamma$ & [-3, 0] & $-1.94^{+0.14}_{-0.12}$ & [-3, 0] & $-1.97^{+0.14}_{-0.12}$ \\
log$_{10}(E_{\rm char, \nu})$ [erg Hz$^{-1}]$ & [30, 36] & $33.45^{+1.07}_{-0.67}$ & [30, 36] & $33.78^{+1.04}_{-0.70}$ \\
\echar [\energyunit] & - & $2.82^{+30.29}_{-2.21}\times 10^{33}$ & - & $6.02^{+60.04}_{-4.82}\times10^{33}$ \\
$f_{\rm Y}$ & [0, 1] & $0.31^{+0.31}_{-0.21}$ & - & - \\
$\tau$ [Gyr] & - & - & (0, 12] &  $1.94^{+1.54}_{-1.31}$
\enddata
\end{deluxetable*}

\section{Results} \label{sec:results}

Fitting the data, we find the energy distribution slope is $\gamma\approx -2$ for both the SFR-SMD hybrid (SSH) model and the constant delay time (CDT) model. $\alpha$ and \echar have wide distributions because they have an inverse correlation that is evident from Eq (\ref{eq:Eflu_rel}), and can be seen in Figure \ref{fig:cornerplot_param_agEf}. Previous findings of $\alpha$ and \echar (e.g. \citealt{Macquart2019, Shin2023, Arcus2024}) have values not consistent within $1\sigma$ of our results. 
For the SSH fiducial model, the fraction of FRB sources tracking the cosmic star formation rate is $0.31^{+0.31}_{-0.21}$, and the median characteristic delay time for the CDT model is $1.94^{+1.54}_{-1.31}$ Gyr. If there are indeed two channels of FRB source formation with independent distributions of delay times from star formation, then both the data and the CDT model are inadequate to capture such a bimodality in delay times. The hypothesis of a purely SFR-tracking population is rejected with $>2\sigma$ confidence for the SSH model, but not for the CDT model. 
However, single parameter delay time distributions are explored in Appendix \ref{sec:delay_time}, which show that the most likely median of the underlying delay time distribution may be close to $1.3$ Gyr, which corresponds well with the monotonically decreasing posterior for $f_{\rm Y}$ with a median $\approx 0.3$. 

In this analysis, the volumetric rate of FRBs $\Phi_0$ has not been constrained as an independent parameter due to the nature of Eq. (\ref{eq:condprob}), which cancels out the dependence on $\Phi_0$. We define the redshift-limited observed all-sky rate $\mathcal{R}_{\rm sky}^{\rm obs}(z\leq 8) = P({\rm SNR} \geq 12) \equiv \Phi_0 K(z\leq 8)$, where we have just taken out the factor $\Phi_0$ for explicit visibility, which can be obtained as follows,
\begin{equation}
    \Phi_{0} = \frac{94}{\Delta t_{\rm base} K(z \leq 8)}.
    \label{eq:phi0}
\end{equation}
In this equation, $\Delta t_{\rm base} = 65.8 \;{\rm days} = 0.18$ yr is the estimated survey duration for the baseband catalog. The process of estimating $\Delta t_{\rm base}$ is discussed in Appendix \ref{sec:base_survey_duration}. 
For the SSH model, $\Phi_0^{\rm SSH} = 5.4^{+1.6}_{-1.3}\times 10^{-5}$ Mpc$^{-3}$ yr$^{-1}$ above the pivot specific energy of $10^{30}$ erg Hz$^{-1}$ at 600 MHz in the comoving frame, where the central value is determined from the 50th percentile of the parameter posteriors, and the error comes from $10^4$ random samples taken from the MCMC chain. Similarly, for the CDT model, we obtain $\Phi_0^{\rm CDT} = 2.8^{+1.5}_{-1.1}\times 10^{-5}$ Mpc$^{-3}$ yr$^{-1}$. To convert from the usage of spectral energy to total energy across the band, for comparison with \citet{Shin2023}, we multiply the spectral energy by their assumed frequency bandwidth of 1 GHz. Thus, the obtained rate can also be interpreted as $\Phi_0^{\rm SSH} = 5.4\times 10^4$ Gpc$^{-3}$ yr$^{-1}$ above $10^{39}$ erg,
% which is roughly 2 orders of magnitude higher than 
which is consistent with the rate inferred by \citet{Shin2023}. 

\begin{figure*}
    \epsscale{1.1}
    \plottwo{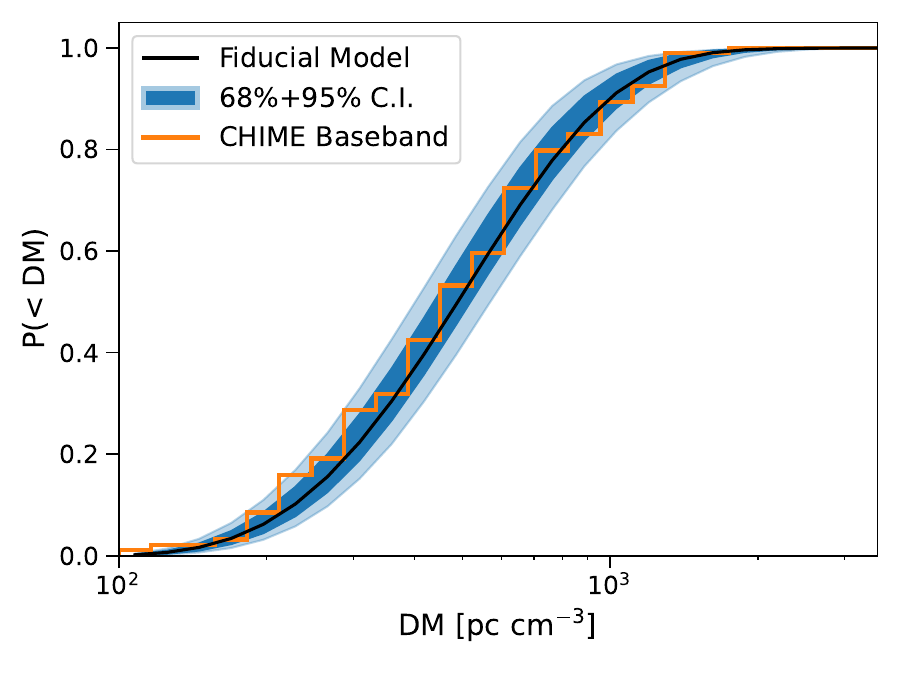}{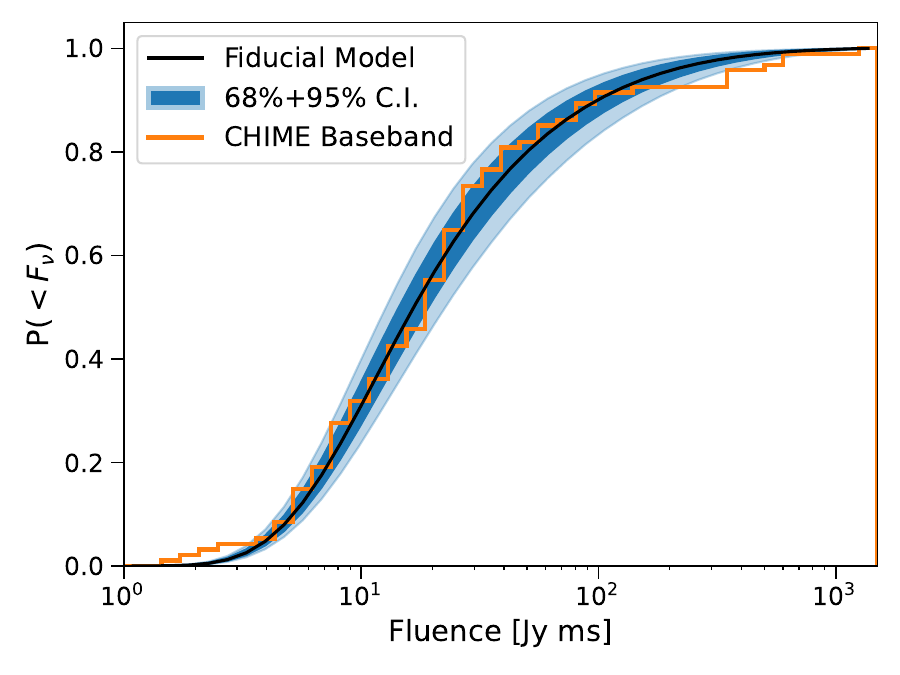}
    \caption{Plots of the cumulative FRB (\textit{left}) DM, and, (\textit{right}) fluence distribution observable at a fluence threshold of 1 Jy ms for the SFR-SMD hybrid model. Both plots also show the cumulative distribution of FRB DMs and fluences in our observed sample (solid orange line).
    }
    \label{fig:cum_DMnFdist}
\end{figure*}

\begin{figure*}
    \gridline{\fig{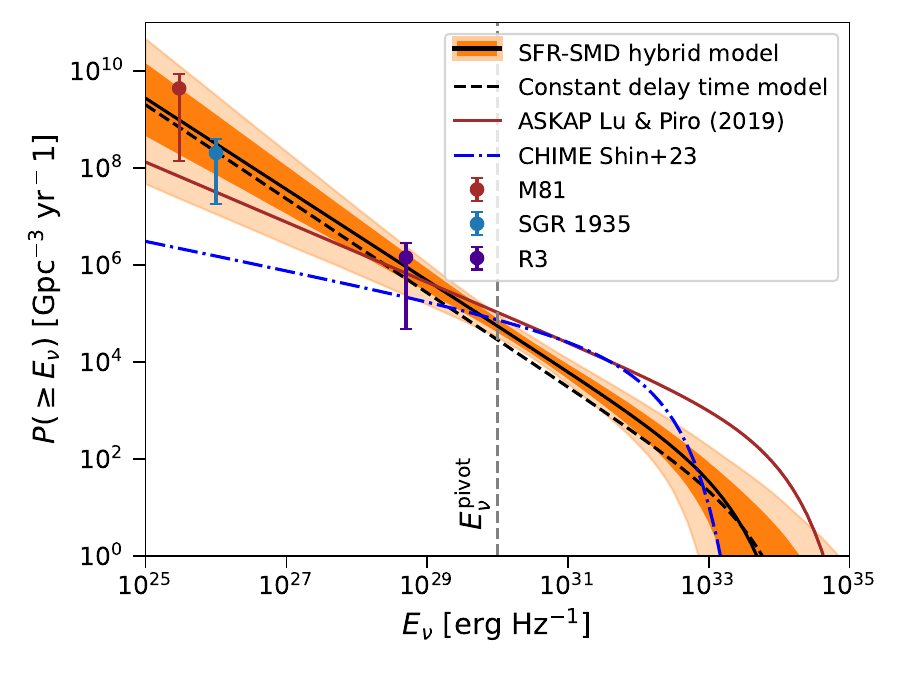}{0.45\textwidth}{(a)} \fig{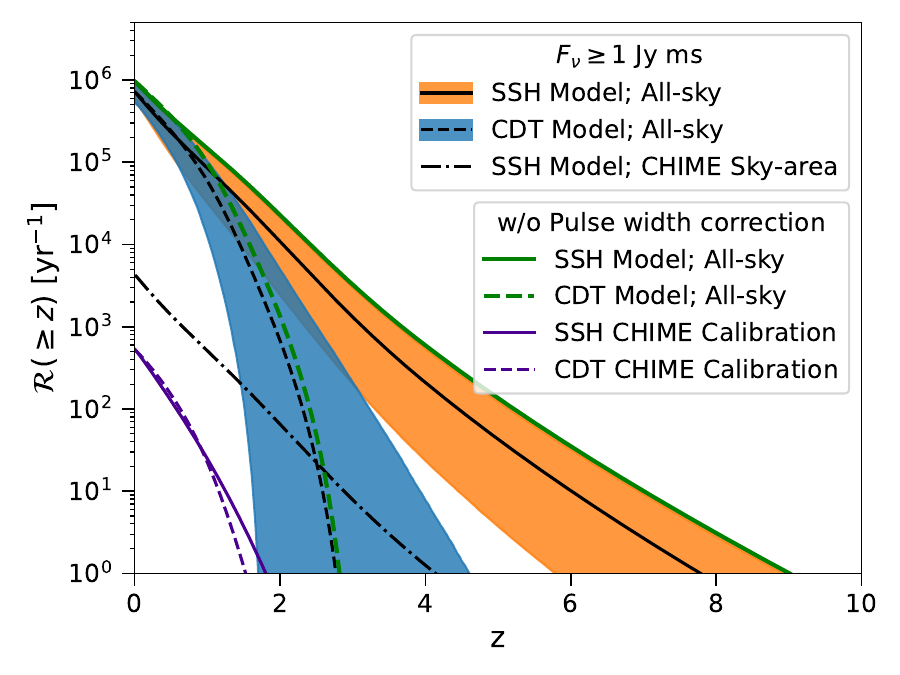}{0.45\textwidth}{(b)}}
    \caption{(a) This figure shows the median (solid black line) and the error (shaded regions) of the cumulative energy distribution determined using the SFR-SMD hybrid model. Comparisons are made with the constant delay time model (dashed black line) and previously determined distributions using data from CHIME \citep{Shin2023} and ASKAP \citep{Lu2019}. Rate estimates for fairly close-by repeaters are also used as markers for comparison at low energies \citep{Lu2022}.
    (b) Cumulative redshift distribution of FRBs observable over the entire sky with fluence thresholds of 1 Jy ms (median lines as indicated with the 68\% confidence interval as a shaded region) for both models.
    % , assuming that the energy distribution extends to low energies with the same power law slope as in the left figure.
    These have been corrected for the effect of an increasing fluence threshold from cosmological pulse width broadening (described in Appendix \ref{sec:app_pulse_width}), and are compared with distributions without this effect accounted for (green lines). The best-calibrated distributions with CHIME selection effects and sky coverage applied to represent the evolution of the observed population are also shown as indigo lines. The effect of telescope selection bias for the SSH model is evident when the black dashed-dotted line using a fluence threshold and CHIME's primary beam sky coverage ($\approx 250$ sq. deg.) is compared to the CHIME observation function calibrated evolution (indigo).
    }
    \label{fig:bestfit_zEdist}
\end{figure*}

Using $10^4$ random samples of the parameters in the MCMC chain, we also generate the cumulative distributions of fluence and DM, which are compared to their observed distributions in Figure \ref{fig:cum_DMnFdist}. To obtain the redshift evolution of the FRB rate from $z=0-14$, $\mathcal{R}_{\rm sky}(z)$, we integrate Eq. (\ref{eq:com_frb_rate}) over all-sky and specific energies above the CHIME threshold fluence $F_{\nu_0}^{\rm th} = 1$ Jy ms. It must be pointed out that $\nu$ in Eq. (\ref{eq:com_frb_rate}) now refers to the comoving frequency corresponding to the observing frequency. This means that volumetric rate of FRBs at $\nu=\nu_{\rm com}$ is also different from 600 MHz as $\Phi = \Phi_0 (\nu_{\rm com}/600 {\rm \;MHz})^{\alpha}$, such that,
\begin{equation}
    \mathcal{R}_{\rm sky}(z) = \frac{dN(z, >F_{\nu_0})}{dt dz} = \Phi \psi_{\ast}(z)\frac{4\pi d_{\rm com}^2(z)c}{(1+z)H(z)} \int_{E_{\nu_{\rm com}}} P(E_{\nu}) dE_{\nu}
    \label{eq:pred_skyrate}
\end{equation}
In addition, the redshift evolution needs to incorporate the effect of temporal broadening of bursts due to cosmological time dilation, and the process is outlined in Appendix \ref{sec:app_pulse_width}. The cumulative redshift evolution is shown in Figure \ref{fig:bestfit_zEdist}(b), showcasing the difference introduced by incorporating pulse width broadening effects. The comparison of the intrinsic redshift distribution with the CHIME calibrated distribution illustrates why bursts which should also be detectable from high redshifts, when considering just the fluence threshold, are not observed due to selection effects.

Since the analysis of the Baseband catalog in this work has been carried out without intrinsically including pulse widths, the results are expected to capture the range of pulse widths observed by CHIME at low redshifts automatically. 
Therefore, we quote the SSH all-sky FRB rate above 5 Jy ms at an observing frequency of 600 MHz with scattering timescales $\lesssim 10$ ms and not additionally corrected for pulse width broadening to be $552^{+47}_{-52}$ day$^{-1}$. The corresponding CDT all-sky rate is $538^{+54}_{-63}$ day$^{-1}$. These values seem consistent with the rate predicted in \citet{CHIME2021} of $525\pm30 ^{+140}_{-130}$ updated by its Erratum, but it is difficult to interpret if this is just a coincidence or not. 
This is because several factors contrast between this analysis and Catalog 1, such as the difference in $\alpha$ and $\gamma$, as well as the systematic difference in fluences between Catalog 1 and its baseband update. In addition, the Catalog 1 all-sky rate has been calculated using an injections-calibrated mapping from fluence to SNR. We note that the fluences and SNRs have a large scatter in both the injections sample (Figure 24 in \citealt{CHIME2021}) as well as the baseband catalog, the latter of which is shown in Figure \ref{fig:scatter_dm_fluence}(b) (in Appendix \ref{app:corr_baseband}).

\subsection{Future Predictions and Implications for Probing Astrophysics} \label{sec:discussion_future}

We show the marked difference, in Figure \ref{fig:bestfit_zEdist}, between the FRB rates obtained for CHIME when the joint DM-$F_{\nu_0}$ selection function is implemented and when it is ignored with just the threshold fluence and CHIME's sky coverage being considered. For future telescopes, it is not known apriori what the exact selection functions would be, so the detection criterion is just assumed to be $10\sigma$. 
Following \citet{Connor2023}, we define the instantaneous sky coverage of a telescope comprised of parabolic reflectors as,
\begin{equation}
    f_{\rm sky} = \frac{1.13}{4\pi} \times \Big(\frac{c}{\nu_0 D}\Big)^2,
    \label{eq:fsky_simple}
\end{equation}
where, $D$ is the diameter of the dish. For telescopes with steerable dishes that have commensal FRB search pipelines, which cannot use the entire baseline, such as DSA-2000 and the mid-frequency component of the first iteration of the Square Kilometer Array (SKA1-MID),
\begin{equation}
    f_{\rm sky} = \frac{1}{4\pi} \min\Big\{ 1.13 \times \Big(\frac{c}{\nu_0 D}\Big)^2, N_{\rm beam} \times 1.13 \times \Big(\frac{c}{\nu_0 d_{\rm bl}}\Big)^2 \Big\},
\end{equation}
where, $d_{\rm bl}$ is the baseline used for FRB search, and $N_{\rm beam}$ is the number of synthesized searchable beams. Then, we define the detection threshold using the sensitivity $\sigma$ of the telescope as follows,
\begin{equation}
    T = 10 \sigma = 10 \times \frac{{\rm SEFD}}{\eta_{\rm c} \sqrt{n_{\rm pol} \Delta \nu \Delta t_{\rm obs}}},
\end{equation}
where, $\rm SEFD$ is the system-equivalent flux density, $\eta_{\rm c}=0.9$ is assumed as the correlator efficiency, $n_{\rm pol}$ is the number of polarizations, $\Delta \nu$ is the bandwidth, and $\Delta t_{\rm obs}$ is the signal integration time. For this subsection, we take $\Delta \nu = 300$ MHz, and $\Delta t_{\rm obs} = 1$ ms. $n_{\rm pol} = 2$ for all listed telescopes, except BURSTT-2048, which has single polarization antennas \citep{Lin2022_burstt}. For SKA1-MID, we have assumed a baseline of 2km for the commensal FRB search pipeline, which optimizes the survey speed and which seems to be the primary mode that the pipeline will operate in. SKA1-LOW uses aperture arrays for which the factor of 1.13 in Eq. (\ref{eq:fsky_simple}) is replaced with 1.48, as for the standard Rayleigh criterion. We also consider each SKA1-LOW station to act like a dish with $D=35$ m \citep{Braun2019_SKA1}.

All planned next-generation telescope rate predictions are made at 600 MHz, with the exception of DSA-2000 and SKA1-LOW, whose values are reported at 850 MHz and 200 MHz respectively. For ultra-wideband telescopes with roughly constant SEFD, such as CHORD and DSA-2000, FRB rates will be smaller at higher frequencies as $f_{\rm sky} \propto \nu^{-2}$. The telescope-specific values for next-gen facilities are noted in Table \ref{tab:tel_params}. We also include specifications for a future telescope, referenced to in the text as OpTel, which is optimized to probe the Epoch of Reionization (EoR). The specifications for OpTel have been chosen to be at 200 MHz because observed rate increases at lower frequencies due to the statistical SED (see Eq. \ref{eq:pred_skyrate}), and also because observing at 200 MHz keeps the rest-frame frequency from becoming too large.
% The specifications for OpTel have been chosen to be at 600 MHz, but the same redshift evolution should be observed for any frequency, given the same values of $f_{\rm sky}$ and threshold $T$. A telescope for ideally observing high-redshift FRBs should have observing bands at low frequencies, such as 200 MHz.

\begin{deluxetable*}{cccccccc} \label{tab:tel_params}
\tablecaption{Next-generation and EoR-optimized radio telescope specifications for FRB search configurations.}
\tablehead{\colhead{Telescope} & $\nu_0$ (MHz) & \colhead{D (m)} & \colhead{$N_{\rm beam}$} & \colhead{Baseline (m)} & \colhead{SEFD (Jy)} & \colhead{$f_{\rm sky}$} &  \colhead{$10\sigma$ (Jy)}}
\startdata
CHORD & 600 & 6 & 512 & - & 9 & $6.24\times 10^{-4}$ & $1.29\times 10^{-1}$ \\
DSA-2000 & 850 & 5 & $10^6$ & 5000 & 5 & $4.47 \times 10^{-4}$ & $8.61 \times 10^{-2}$ \\
BURSTT-2048 & 600 & - & 2048 & - & 600 & $0.24$ & $12.17$ \\
SKA1-MID & 600 & $15$ & $4.5\times 10^5$ & 2000 & 5 & $9.98\times 10^{-5}$ & $2.44\times 10^{-2}$ \\
SKA1-LOW & 200 & 35 & $5\times 10^4$ & 2000 & 9 & $2.17\times 10^{-4}$ & $1.29\times 10^{-1}$ \\
OpTel & 200 & - & - & - & 0.07 & $0.01$ & $10^{-3}$
\enddata
\tablecomments{Telescope properties computed with 300 MHz bandwidth and 1 ms integration time. $\nu_0$ mentioned above may not be the central frequency of the telescope.}
\tablerefs{\citet{Vanderlinde2019}, \citet{Hallinan2019}, \citet{Connor2023}, \citet{Lin2022_burstt}, \citet{Braun2019_SKA1}}
\end{deluxetable*}

\begin{figure*}
    \epsscale{0.6}
    \plotone{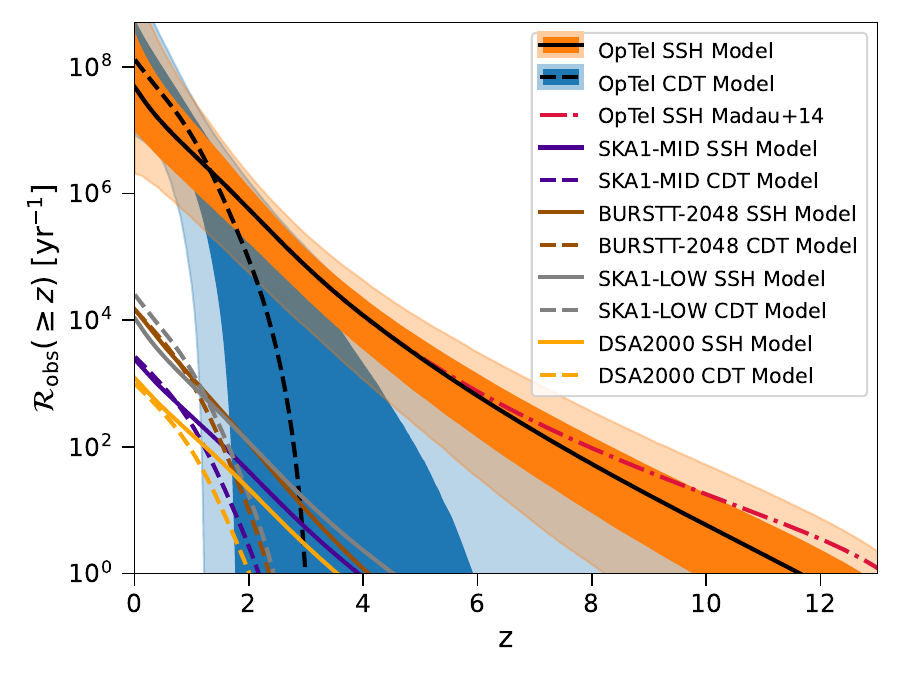}
    \caption{Predictions for the cumulative observed redshift distribution of next-generation radio telescopes, with forecasts included for both the SFR-SMD hybrid model (solid lines) and the constant delay time model (dashed lines). The detection threshold is assumed to be a signal-to-noise ratio of 10, with cosmological pulse width broadening effects included (as described in Appendix \ref{sec:app_pulse_width}). Error on rates are not depicted for planned telescopes, which are provided only for the hypothetical telescope, OpTel (shaded regions show 68\% and 95\% C.I.), which has a system-equivalent flux density of 0.07 Jy and a sky coverage of $\gtrsim 400$ square degrees. 
    The median SFR-SMD hybrid model rate computed with the cosmic star formation rate density history estimated in this work using JWST results (solid black line) is compared to the pre-JWST determination from \citet{Madau2014} (crimson dashed-dotted line). }
    \label{fig:zpred}
\end{figure*}

The predictions for the observed redshift distribution of FRBs with these next-generation telescopes are depicted in Figure \ref{fig:zpred}, with OpTel predictions showing the large range of possible redshift evolutionary tracks because current data are not sufficient to exactly constrain them beyond $z\gtrsim 1$. Models more realistic than a constant delay time distribution will extend to higher redshifts, consequently confining the redshift evolution between the two evolutionary tracks shown. The redshift evolution of FRBs observed by SKA1-MID is very similar to that seen by CHORD, which is why the redshift distribution for the latter has not been shown separately in Figure \ref{fig:zpred}. Notably, all next generation telescopes do not seem to have the capability to detect FRBs in the EoR. Finally, it is observed that the redshift distribution that would be seen by OpTel differs by an order of magnitude at $z\gtrsim 12$ between the redshift evolution determined using JWST and that determined previously by \citet{Madau2014}.

Since, CHORD, DSA-2000 and SKA1-MID will localize FRBs to sub-arcsecond precision, they will be very useful in finding and characterizing host galaxies, with redshift information from spectroscopic surveys such as the all-sky survey SPHEREx \citep{Alibay2023_spherex}. In addition to a variety of proposed science with the observed redshift distribution, it would be much easier to constrain the delay time of FRB progenitor formation.
If FRBs are revealed to follow large delay times with tight distributions, then the future of high redshift FRB science seems grim.

To use FRBs as a probe of the EoR, most studies utilizing FRBs without redshift information require thousands, if not tens of thousands of DM measurements for FRBs in the EoR (\citealt{Beniamini2021}, \citealt{Pagano2021}, \citealt{Dai2021}, \citealt{Shaw2024}). Similar requirements also exist for constraining the epoch of second helium reionization around $z=3-4$ \citep{Bhattacharya2021, Caleb2019}. We argue a need for the planning and development of telescope capable of answering these fundamental questions through FRBs, like the hypothetical OpTel. OpTel would be able to detect $5.8^{+4.6}_{-3.9}\times 10^4$ FRBs per year at $3<z<4$ according to the SSH model. This number is sufficient to resolve the epoch of second helium reionization with good confidence, even without redshifts.

To resolve the mysteries of the EoR, collecting a large number of FRB DMs can ideally be done within a few years, as OpTel can detect $630^{+730}_{-485}$ FRBs per year at $z\gtrsim 6$. 
As next-gen telescopes are poised to localize FRBs with great precision, it is expected that OpTel would have even better localization capability, which will enable the identification of host galaxies and correspondingly, their redshift at high-z.
We use a fraction of $\sim 20\%$ as the rate of identification of an FRB host galaxy in UV/optical follow-up and acquisition of its spectroscopic redshift, following the fraction of redshifts obtained for GRBs \footnote{\url{https://www.mpe.mpg.de/~jcg/grbgen.html}}. With this rate, it will take 5 years with 100\% on-sky time to acquire at least 10 redshift-identified FRBs in each redshift bin of $\Delta z=1$ in $6\leq z \leq 10$. Calculating IGM electron densities in these redshift bins will enable the differentiation between reionization histories that are either too fast or too slow \citep{Beniamini2021}. \citet{Heimersheim2022} predict that 100 localized FRBs would be required to constrain the beginning, mid-point and end of EoR to better than an accuracy of 11\%, which could also be attained within a few years. All of these predictions assume that the instrumental selection effects, other than the ones implemented in Appendix \ref{sec:app_pulse_width}, are non-existent, which in reality would reduce the number of detected events and increase the time for obtaining the same results. However, we have placed strict conditions on the detectability, further suggesting that the threshold for detecting higher redshift bursts is higher. These stipulations, in reality, may become somewhat relaxed. As the next generation of radio telescopes are being built, the community must also deliberate the development of the next-next generation.

\section{Discussion} \label{sec:discussion}
The results yield a clearer picture of intrinsic FRB distributions.
The congruence of energy distribution slopes (see Section \ref{sec:Edist_and_rate}) derived from different telescopes marks a turn towards our ability to probe the intrinsic FRB energy distribution, contrary to previous conclusions (e.g., \citealt{Hackstein2021}). The redshift distribution (Figure \ref{fig:bestfit_zEdist}b) is primarily obtained using the constrained energy distribution, the choice of a model describing the redshift evolution, and the volumetric rate inferred based on both. 
We discuss next the caveats and conditions associated with the results, compare them to results in the literature, outline the reasons for various choices taken within the pipeline, and test the robustness of our findings against certain changes in the analysis or future modifications/updates to the data. Finally, we contemplate the properties of the intrinsic FRB progenitor population, based on current evidence. 

\subsection{Challenges in Determining Star Formation Rate Density at High z}
There are certain issues in the computation of the SFRD that can affect our predictions in the moderate to high redshift range. SFRD estimates are good at low redshifts because the dust correction relation and mass-to-light ratio $\kappa_{\rm UV}$ have been calibrated for local Universe galaxies. This is fortuitous for calibration with data because all FRBs observed till now are at low redshifts. Dust attenuation is expected to decrease as redshift increases, but how the $A_{\rm UV}-\beta$ relation changes with redshift is uncertain. 

With the IMF expected to become more top-heavy, and luminosity being a strongly increasing function of stellar mass, the mass-to-light ratio could decrease with increasing redshift. Consequently, one should infer a lower stellar mass in high-redshift galaxies; indeed, without these considerations, some claims have been made that high-redshift galaxies are overly massive \citep{labbe23}, though likely not to the point where they violate $\Lambda$CDM \citep{chworowsky24}. 
A larger proportion of massive stars, due to a top-heavy IMF, would yield an increasing fraction of likely FRB source progenitors, making our SFRD predictions highly uncertain at $z>10$. 

IMF measurements are restricted to the very local neighbourhood of the Milky Way. Furthermore, it is highly dependent on metallicity, as higher mass stars are expected to form when the star-forming gas was relatively metal-free. But the exact form and evolution of a universal IMF remains a mystery (see e.g. \citealt{Madau2014}, \citealt{Bromm2004}).  Although a Salpeter IMF is disfavored by observations of the local Universe, it has the same slope at masses $> 1$ M$_{\odot}$ as other commonly used IMFs, where the progenitor stars of FRB sources are expected to be. As such, a Salpeter-derived $\kappa_{\rm UV}$ used in this work is appropriate for the mass range that is of interest.

\subsection{Choice of $E_{\nu}^{\rm pivot}$} \label{sec:discussion_Epivot}
With a power-law slope of $\gamma\approx -2$, the number of FRBs predicted at energies below $E_{\rm char, \nu}$ rapidly increases. But the counteracting effect of a threshold fluence and survey incompleteness negatively impact the number of FRBs observed. 
Even though the minimum observed isotropic FRB specific energy is $\sim 10^{21}$ erg Hz$^{-1}$, for SGR 1935+2154 \citep{Giri2023}, suggesting that FRB energies extend at least about 9 orders of magnitude below the pivot energy, FRBs may or may not have a different power-law slope for $E_{\nu} < E_{\nu}^{\rm pivot}$. 

\begin{figure*}
    \gridline{\fig{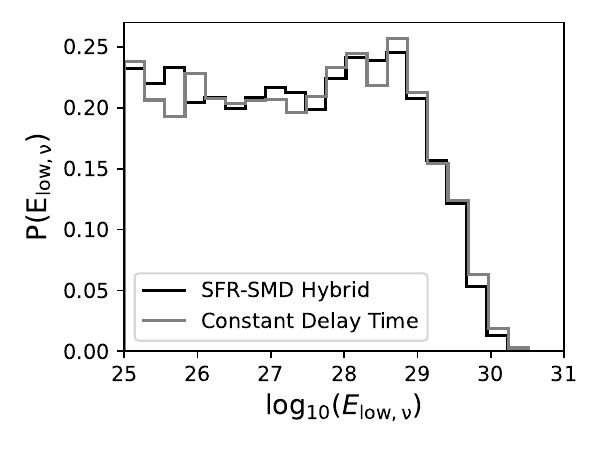}{0.4\textwidth}{(a)}
    \fig{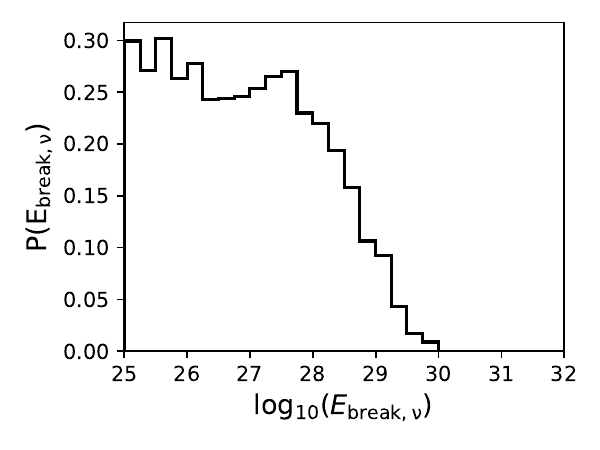}{0.4\textwidth}{(b)}
    }
    \caption{(a) Posterior distribution of the low-end energy cutoff $E_{\rm low, \nu}$, which shows that an abrupt cutoff is allowed in both our fiducial models at $E_{\rm low, \nu} \lesssim 10^{30}$ \energyunit, indicating that the data are not sufficient in this regime to dictate the slope of the obtained energy distribution. (b) Posterior distribution of the spectral energy distribution power-law break for the SFR-SMD fiducial model, $E_{\rm break, \nu}$, below which we have assumed that the power-law steepens to -2.75. This shows that a good chance of such a steepening occurring is below $10^{28}$ \energyunit, where we ascertain only a small fraction of bursts in our sample to be, thereby ruling out such a possibility in the current data.}
    \label{fig:E_low}
\end{figure*}

To test the lowest energy where we can confidently quote all-sky FRB rates with the baseband sample and choose a value of $E_{\nu}^{\rm pivot}$, we postulate that the energy distribution cuts off abruptly at $E_{\rm low, \nu}$, marking the low energy end. Simply put, $P(E_{\nu}) = 0$ for $E_{\nu} < E_{\rm low, \nu}$. 
With the fiducial model parameters for both models, we use an MCMC analysis to determine the posterior of $E_{\rm low, \nu}$, shown in Figure \ref{fig:E_low}a, starting with a uniform prior. An abrupt low-end cutoff is not tolerated by the data through both fiducial models for $E_{\rm low, \nu} > 10^{30}$ \energyunit. Thus, the regime with $E_{\nu} < 10^{30}$ \energyunit $ \equiv E_{\nu}^{\rm pivot}$ cannot be confidently explored using this analysis, and either more data or a more sensitive survey is required to probe these low energy bursts in the nearby Universe.

\begin{figure*}
    \gridline{\fig{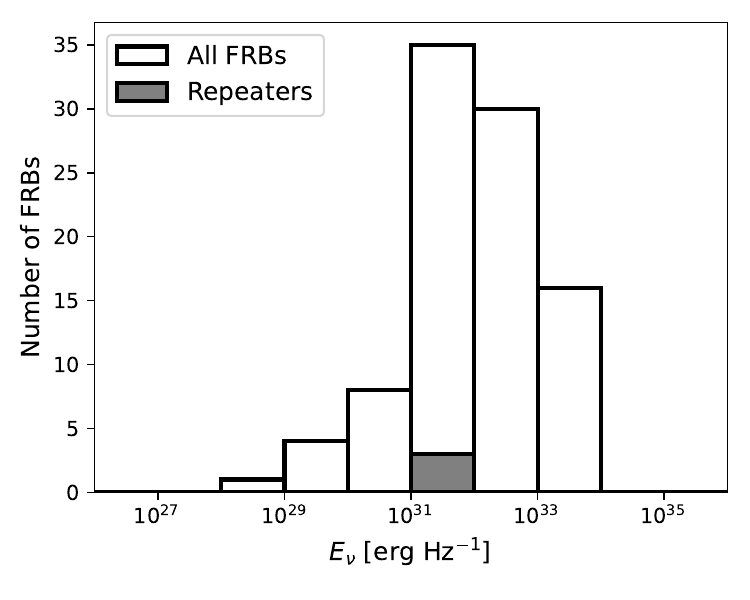}{0.45\textwidth}{(a)} \fig{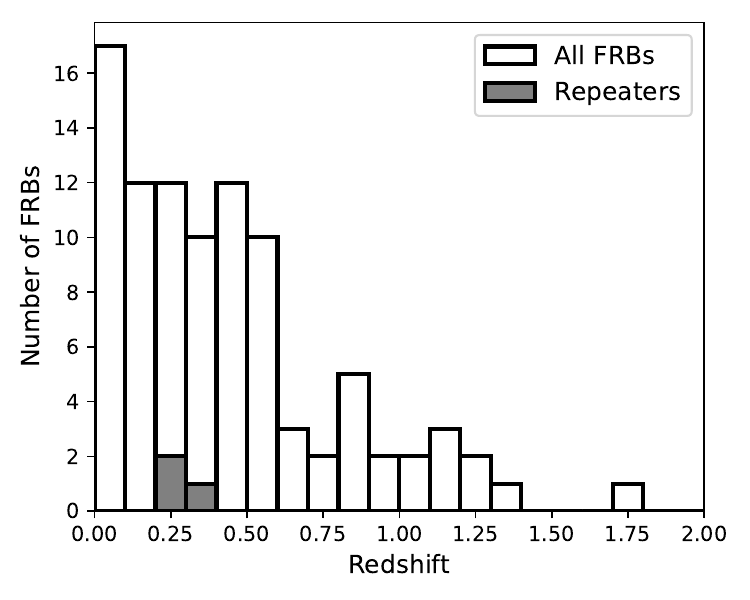}{0.45\textwidth}{(b)}}
    \caption{Distribution of the best-fit (a) energy, and (b) redshift of each of the 94 baseband FRBs in our observed sample. Both figures mark the distinction between 3 bursts from repeating sources.
    }
    \label{fig:ind_zEdist}
\end{figure*}

\subsection{The Energy Distribution and Spectrum} \label{sec:Edist_and_rate}

Both $\gamma$ and $E_{\rm char, \nu}$ obtained in this work are not unprecedented. $\gamma$ is consistent to within $1\sigma$ with values obtained in previous studies, such as $-1.6 \pm 0.3$ \citep{Lu2019}, $-1.79^{+0.31}_{-0.35}$ \citet{Luo2020}, $-2.09^{+0.14}_{-0.10}$ \citet{James2022a}, $\approx -1.9$ \citep{Zhang2022}, and $\in (-1.8, -2.1)$ \citep{Lin2024}. Power-law slopes of some sub-samples of \citet{Arcus2024} may also be consistent. \echar obtained in this work is also consistent with some previous studies \citep{Lu2019, Luo2020, Lin2024}.

A potential reason for \citet{Shin2023} obtaining a flatter slope of $-1.3^{+0.7}_{-0.4}$ may be the large scatter between recorded SNRs and burst fluences. We note from a simple analysis of the baseband catalog that even though there are no high fluence bursts ($F_{\nu_0}>60$ Jy ms) above a DM of 750 pc cm$^{-3}$, there are both low and high SNR bursts at these high DMs (see Figure \ref{fig:scatter_dm_fluence}(b) in Appendix \ref{app:corr_baseband}). As \citet{Shin2023} model the intrinsic DM and fluence, but use SNR to evaluate their model assuming a correlation between fluence and SNR, their MCMC analysis naturally associates bursts with high SNR to high intrinsic energies, which is not necessarily correct. This inflates the fraction of FRBs near $E_{\rm char, \nu}$ decreasing the slope of the energy distribution.

In our pipeline, $\Phi_0$ is a normalizing factor and not an inferred parameter, which makes it straightforward to understand the origin of the volumetric rate (see Eq \ref{eq:phi0}). The volumetric rates inferred in this study are close to those obtained at $\sim 10^{30}$ \energyunit by previous studies \citep{Shin2023, Chen2024, Lu2019, Luo2020}, which is also seen in Figure \ref{fig:bestfit_zEdist}. The reason for its correspondence with \citet{Shin2023} may be because the minimum energy where the power-law can be constrained with confidence, is the same for both this work and \citet{Shin2023} (see Section \ref{sec:discussion_Epivot}).

Next, we discuss the caveats related to the assumption of a power-law SED shape. The maximum rest-frame frequency reached in this work is as high as 7 GHz (in Figure \ref{fig:bestfit_zEdist}b), which has only been seen for a few cases (e.g., \citealt{Gajjar2018}). The k-correction is significant for high-redshift predictions, given the steep value of $\alpha$. Since predictions at, say 200 MHz, require that the power-law extend only up to 3 GHz, OpTel has been designed at 200 MHz to provide results more resistant to bias in parameters.

\subsection{Robustness of Model Parameters} \label{sec:discussion_robust}

One of the biggest unknown factors in the pipeline is the observation function $P({\rm SNR | DM}, F_{\nu_0})$. Due to a limited number of FRBs being injected for creating the observation function, and missing values being interpolated to in our pipeline, the result is tested against changes to the observation function. A linear function in log-log space with fluence is created, with its fluence end points having the same value as the observation function, and which is constant with DM. For both SSH and CDT models, all parameters change significantly ($\gtrsim 1\sigma$), indicating that the result is sensitive to the shape of observation function. 

Furthermore, the observation function has a steep decline at fluences $\lesssim 2$ Jy ms. The slope of the observation function at these fluences strongly dictates the SED index $\alpha$ and \echar. Reducing the slope by raising the value of the observation function at 1 Jy ms by roughly an order of magnitude increases the SED index to $\approx -1.6$, and lowers \echar to $\sim 10^{33}$ \energyunit. Finally, we test whether omitting fluences $< 2$ Jy ms from the observation function and from the data has any effect on the results. Both SSH and CDT models yield results similar to the fiducial parameters. 

Deviations from the current injections-determined observation function arising from a possible future injections campaign are not expected to be significant, but one possibility we have explored above is that the observation function may not be as steep. Even though we have used a statistical SED, the generally accepted value in previous literature is $\alpha=-1.5$ (e.g. \citealt{Macquart2019}). If $\alpha$ is not really as steep, then our median redshift predictions actually represent a conservative estimation. However, a recent estimation of the injections-corrected spectrum in the observer frame using CHIME Catalog 1 data finds $\alpha=-2.29\pm0.29$ ($\alpha=-2.50\pm0.43$ for non-repeaters only; \citealt{Cui2025}). 
% \sout{However, parts of the parameter space that may be underpopulated, such as at low fluences, may be affected. An} \sout{improvement in this regard may help fit the observed fluence distribution in Figure \ref{fig:cum_DMnFdist} better. From observations, $\alpha$ is} \sout{expected to be $\approx -1.5$ based on the literature (e.g. \citealt{Macquart2019}), one possibility we have explored above is} \sout{that the observation function may not be as steep.}

Another big effect may come from the estimation of the  $\rm DM_{IGM}$ distribution, which is mostly informed through simulations \citep{McQuinn2014, Jaroszynski2019, Zhang2021b}. For example, \citet{Zhang2021b} estimate this distribution using the IllustrisTNG simulations at many redshifts, and provide fitting functions for the same. The robustness of our results is tested against an implementation of $\rm DM_{IGM}$ distributions from \citet{Zhang2021b}. The median SSH model parameters become $\alpha=-3.65$, $\gamma=-1.97$, \echar$=10^{33.40}$ \energyunit and $f_{\rm Y}=0.31$, while the median CDT values change to $\alpha=-4.24$, $\gamma=-1.97$, \echar$=10^{33.57}$ \energyunit and $\tau=2.72$ Gyr. Most conspicuous are the changes in $\alpha$ for both models, and in $\tau$ for CDT, even though these changes correspond to $<1\sigma$ deviation from the fiducial values.

We also establish the $\sim 10\%$ errors in baseband fluences \citep{CHIME2023} do not affect parameter posteriors. Finally, we note that $\alpha$, $\gamma$ and \echar for delay time distribution implemented in Appendix \ref{sec:delay_time} remain the same as as for the CDT model.

Our constrained parameters may be biased due to propagation-induced effects \citep{Bhardwaj2024b}. Assuming that the energetics of FRBs and their electron column density accrued through their sightline in their respective host galaxies are not correlated, there is a larger bias against detecting dimmer bursts traversing long host galaxy sightlines than against detecting bright bursts. If such a bias exists, the intrinsic energy distribution slope would be steeper. 

\subsection{Repeaters vs (Apparent) Non-repeaters} \label{sec:discussion_repeat_vs_nonrepeat}

In general, CHIME will not probe low luminosity bursts from repeaters, unless they are very close by, and selecting a burst with $E_{\nu} > $ \echar in a random sample from the same source has a very low probability. This is exactly what is observed in Figure \ref{fig:ind_zEdist}(a), where the gray boxes give the location of the 3 repeaters present in our analysis, showing that the selected bursts are deduced to have ordinary energies within the CHIME sample. As a consistency check, it is noted that the most likely redshift of these repeaters, shown in Figure \ref{fig:ind_zEdist}(b), is less than their maximum estimated redshift based on their DM, with $z_{\rm max} \sim 0.5$ \citep{Michilli2023}.
Removing repeaters has a negligible effect on the determination of the energy and redshift distribution parameters, which may simply reflect the small fraction  of repeaters in the sample ($\approx 3\%$).

Two hyperactive repeaters have been shown to have a flatter energy distribution slope than we have obtained. For example, FRB 20220912A has a Schechter function power-law slope of $-1.12^{+0.17}_{-0.13}$ above $3.2\times 10^{30}$ \energyunit \citep{Ould-Boukattine2024}. It is observed that the burst rate energy distribution steepens below this energy, which is a property also exhibited by another hyperactive repeaters FRB 20201124A, for which the break occurs at $\sim 8\times 10^{30}$ \energyunit \citep{Kirsten2024}. The energy distribution slope of FRB 20220912A, if it is representative of hyperactive repeaters, seems to challenge the notion that one-off FRBs and repeaters follow the same energy distribution. 

It is also possible that the FRB population follows a broken-power law, but CHIME does not have sufficient sensitivity to resolve the break. To probe this possibility, we perform an MCMC analysis to constrain $\alpha$, $\gamma$, the break energy $E_{\rm break, \nu}$, and the fraction of SFR-following FRB sources $f_{\rm Y}$, within the scope of the SSH model. In this particular case, $\gamma$ defines the slope of the high energy part, while the steeper low energy slope is fixed to $-2.75$ \citep{Ould-Boukattine2024}, assuming that the observed cumulative energy distribution of burst rates is interchangeable with the cumulative energy distribution. $\alpha$, $\gamma$ and $f_{\rm Y}$ remain at their nominal values, and from Figure \ref{fig:E_low}(b) we see that $E_{\rm break, \nu}$ is non-negligible only below $10^{30}$ \energyunit, and roughly flattens only below $10^{28}$ \energyunit. This indicates that even on providing sufficient freedom to yield a plausible break energy, with a uniform prior between $10^{25}$ and $10^{33}$ \energyunit, a break cannot be confidently deduced within the current dataset. Moreover, the observed repeater breaks are $> 10^{30}$ \energyunit, which is ruled out at $>2\sigma$ by the MCMC analysis. Future data with more bursts at lower energies may provide more concrete results, but for now, it seems that the statistical energy distribution of typical apparent non-repeaters is different from that of hyperactive repeaters, such as FRB 20220912A and 20201124A.

\subsection{A Single Population of FRB Emitters Dominates the Energy Distribution}
Even though FRB progenitors may be formed from different channels, the population of FRB sources dominating the energy statistics seems to be the same within $10^{30} < E_{\nu} < 10^{34}$ \energyunit\ because FRBs in this range are described by a single power-law model with exponential cutoff (the cutoff is assumed). One naturally expects to find lower rates at higher energies, but if there is more than one dominant population of FRBs, the power-law slope could steepen at a certain lower spectral energy, much like the cosmic ray spectrum. Another reason may be that such a possible steepening is intrinsic to FRB sources, but even this can be evaded if the break is a property only of hyperactive repeaters comprising a very small fraction of the population (discussed in Section \ref{sec:discussion_repeat_vs_nonrepeat}), or if all sources exhibit a break in their individual energy distributions but at different energies. In the latter scenario, the envelope of the individual energy distributions of all sources may mimic a single power-law. 

Another possibility is that the break occurs at smaller energies, hidden from the scope of current sensitivities. 
However, the correspondence of three localized repeater rates from \citet{Lu2022} with the constrained power law in Figure \ref{fig:bestfit_zEdist}a, probing much lower energies than $E_{\nu}^{\rm pivot}$, suggests that the same power law may well extend down to spectral energies of $\sim 10^{25}$ \energyunit. At this energy, the volumetric rate could be $\approx 2.74^{+10.57}_{-2.25}\times 10^{9}$ \gvolrate. A confirmation of this hypothesis may have to wait for a larger local Universe sample of FRBs, but if this is the case, then repeaters are a part of this single population of FRB sources emitting FRBs over 8 orders of magnitude in spectral energy. 
% It is difficult to explain naturally two distinct populations either having the same power-law slope, or populating different parts of the energy range without having a break or gap in between.

With the assumption of a single continuous power-law slope below \echar, we can rule out the FRB rate being dominated by catastrophic events such binary mergers of compact objects ($< 10^3$ \gvolrate; \citealt{Santoliquido2021}) or long GRBs ($79^{+57}_{-33}$ \gvolrate; \citealt{Ghirlanda2022}). However, just from a volumetric rate standpoint, we cannot rule out core-collapse supernovae as being the sole progenitor channel of the underlying FRB source population, because they have a rate $\sim 10^5$ \gvolrate \citep{Melinder2012, Perley2020}. Furthermore, neutron stars born may have activity periods lasting upto $10^6$ years \citep{Beniamini2023}, which allows even a small fraction of the resulting compact objects to explain the FRB rate above $10^{30}$ \energyunit. A large distribution of rates per source can partially be attributed to a distribution in the relative inclination between the magnetic and spin axis \citep{BeniaminiKumar2024}, which could result in a large range of repetitions even if different sources were otherwise identical.

\subsection{Clues Regarding the Delay Times of Fast Radio Burst Sources}

A redshift-dependent functional form calibrating the evolution of FRBs with cosmic time can only provide clues about what the underlying source population may be. This is because the evolution of FRBs can be parametrized in various ways and with many distinct functions. The data indicates that FRBs may not be tracking solely SFR, unlike the conclusions made by \citet{Shin2023} and \citet{James2022a}. It also may not fully follow SMD, unlike the claim of \citet{Hashimoto2022}, nor does it seem to follow very large delay times, unlike the conclusions of \citet{Zhang2022}. However, hybrid models seem more likely, with delays $\lesssim 3$ Gyr from star formation.

Recent studies of FRB host galaxies \citep{Sharma2024, Bhardwaj2024a, Gordon2023, Law2024} show that most host galaxies are star-forming and massive. 
However, the existence of the globular cluster FRB 20200120E in a star-forming galaxy suggests that progenitors formed from old systems may reside even in star-forming galaxies, and information about more local environments is required to form conclusions. Neutron stars formed from binary white dwarf coalescences in globular clusters seem to contribute at most 1\% to the total FRB rate \citep{Rao2024}. If the obtained delays are real, this hints at the possibility of FRB progenitors forming in the discs of galaxies from the core-collapse of merger remnants \citep{Sharma2024}, which could provide some delay from star formation, along with non-CCSN formation channels, such as accretion- or merger-induced collapse of massive white dwarfs or binary neutron star mergers. Upcoming work (Bhardwaj et al. in prep.) proposing delay times up to a Gyr with electron capture supernova occurring in binary systems may be an additional avenue to achieve the average delay times being found in the data.

Furthermore, the low number of quiescent spiral hosts and a single candidate elliptical galaxy \citep{Sharma2024} among $\mathcal{O}(90)$ host galaxies in the literature can be explained by median delay times $\sim 1.3$ Gyr. The most massive ellipticals ($\gtrsim 10^{11}$ M$_{\odot}$) are mostly already in place at $z\sim 1$, and most of the evolution in terms of transition from the population of spiral to elliptical galaxies takes place between $z\approx0.3-1$ for stellar mass $< 10^{11}$ M$_{\odot}$ (e.g. \citealt{Bundy2005, Oesch2010}). For both the most massive ellipticals, which are ``red and dead'' at $z\approx 1$, and ellipticals which formed later, we can place a conservative upper limit on delay times of roughly 6 Gyr, which rules out large delay times of 10 Gyr or more determined by \citet{Zhang2022}. For the delay times we have determined, elliptical galaxies in the nearby Universe in general, will not host FRBs, with our hypothesis being that the FRB progenitors formed and also died within the age of these ellipticals (or time elapsed since last major merger). This scenario is, however, very distinct from FRB sources being formed in highly dynamic and interacting environments, such as globular clusters on the outskirts of elliptical galaxies \citep{Eftekhari2024, Shah2024}. A consequence of this hypothesis is that the fraction of elliptical hosts will increase at an epoch where they had active star formation in recent history. 

\section{Conclusions} \label{sec:conclusions}
In this study,  we have provided constraints on the energy and redshift distributions of FRBs using joint distribution of the Dispersion Measure and fluence recorded by the CHIME/FRB experiment, which is a superior method than assuming independent distributions for these FRB observable and has only been partly used in one study before this \citep{Shin2023}. More importantly, we use accurate fluence information for the first time, utilizing the CHIME/FRB Baseband Catalog. 

We obtain a model-independent power-law slope of the energy distribution of FRBs of $-1.94^{+0.14}_{-0.12}$, and an exponential cutoff that occurs around $5\times 10^{33}$ \energyunit. A volumetric rate of FRBs of $\sim 5\times 10^4$ \gvolrate above an isotropic-equivalent energy of $10^{30}$ \energyunit $\;$ is obtained, which we also constrain to be the energy above which we are confident about our energy distribution determination from the CHIME/FRB data. Our volumetric rate is consistent with previous deductions \citep{Shin2023, Chen2024, Lu2019, Luo2020}. 
We obtain an all-sky rate of FRBs of about 552 FRBs per day above a fluence threshold of 5 Jy ms at 600 MHz with scattering time $\lesssim 10$ ms.

We also investigate the delay times of FRB sources relative to cosmic star formation using complementary models. It is shown with a constant delay time model that FRBs are likely dominated by sources forming with delays ranging from 0 to 3 billion years, with constant delay time model suggesting delays of $1.94^{+1.54}_{-1.31}$ Gyr. Assuming certain forms of the distribution of delay times (in Appendix \ref{sec:delay_time}), hints of a median delay of $\sim 1.3$ Gyr are identified. However, the possibility of a purely star formation tracking source population can be discarded at $>2\sigma$ confidence for only the SFR-SMD hybrid model, which predicts that only $31^{+31}_{-21}$\% of FRB sources may be associated with cosmic star formation. An interpretation of this result is that there may be two (or more) progenitor formation channels, each producing FRB sources with a distinct distribution of delay times from star formation. Such an interpretation is consistent with a small, but growing evidence of sources localized to distinct environments, corresponding to both young and delayed formation channels. It is possible that magnetars can be formed via certain delayed channels, and that such channels may be more common than previously thought.

We find that the probability of detecting FRBs at $z>5$ with next-generational radio facilities is low, with a chance of around 1 burst at $z\sim5$ collectively between all planned telescopes listed in Table \ref{tab:tel_params}.
We suggest developing a telescope having an instantaneous field-of-view of $\gtrsim 400$ sq. deg. and a system-equivalent flux density of $\leq 0.07$ Jy (bandwidth of 300 MHz and integration time of 1 ms). This corresponds to a $10\sigma$ fluence detection threshold of $\approx 1$ mJy ms.
According to our calculations, these specifications represent the requirements for a telescope capable of constraining the Epoch of Reionization, with detection rates of $630^{+730}_{-485}$ per year at $z \gtrsim 6$ and $53^{+83}_{-43}$ per year at $z\gtrsim 8$.

Finally, it is important to note that CHIME may be missing a significant number FRBs with scattering times greater than 10 ms \citep{CHIME2021}. With only a handful of bursts with high scattering times, the fiducial model in \citet{CHIME2021} remains poorly constrained. These bursts are not accounted for in the observation function, and therefore, more bursts than are estimated may exist. With CHIME Catalog 2 and a larger injections campaign, it may become possible to perform a scattering time analysis simultaneously with a DM-fluence analysis.

\begin{acknowledgments}
OG acknowledges and thanks Kiyoshi Masui for providing CHIME data products and many helpful insights implemented in this study. OG also thanks Kaitlyn Shin, Mohit Bhardwaj, and Ziggy Pleunis for enlightening discussions, and Robert Braun and Pragya Chawla for useful comments regarding SKA and CHIME, respectively. OG is indebted to Jahnavi Malagavalli for assisting the setup of a parallelizable code and providing supercomputer-related suggestions. OG also thanks Maximilien Franco and Hollis Akins for discussions and/or comments regarding the high-redshift SFR estimation pipeline. The authors thank the referee for suggestions which significantly improved the clarity and quality of this work.

The authors acknowledge the Texas Advanced Computing Center (TACC) at The University of Texas at Austin for providing HPC resources through project AST22018. The authors also acknowledge that UT Austin, where this research has been conducted, is on the Indigenous land of the Tonkawa people, and historically, the Comanche and Apache moved through this area.

The work was funded in part by an NSF grant AST-2009619 (PK and OG), a NASA grant 80NSSC24K0770 (PK, PB and OG), a grant (no. 2020747) from the United States-Israel Binational Science Foundation (BSF), Jerusalem, Israel (PB) and by a grant (no. 1649/23) from the Israel Science Foundation (PB).

Baseband burst selection code, most data products generated/used in this study, and code to visualize them are available on GitHub at: \url{https://github.com/omguptaup/frb-cosmic-evol}.
\end{acknowledgments}

\software{FRB \citep{FRB:2023}, astropy \citep{astropy:2013, astropy:2018, astropy:2022}, scipy (\url{https://scipy.org/}), emcee \citep{Foreman-Mackey2013}, corner \citep{corner:2016}, dynesty \citep{dynesty:2020}}

\appendix 

\section{Dust correction to specific UV luminosity density} \label{sec:dustcorr}
Dust attenuates more at higher frequencies, absorbing UV and optical light, re-radiating it in the infrared. Consequently, dust causes the blue UV-continuum slope $\beta$ of a galaxy's spectrum to become redder. While $\beta$ is also sensitive to the age and metallicity of a galaxy, it is particularly sensitive to the dust content. It has been seen that there is an inverse correlation between $\beta$ and the absolute magnitude $M_{UV}$, and we use the straight line fits for the $M_{UV}-\beta$ relationship found in \citet{Bouwens2014}, at various redshifts, to find $\beta$ for all $M_{UV}$ bins defining the UVLF. We use the linear fit from \citet{Takeuchi2012}, updated from the widely used fit found in \citet{Meurer1999}, to find the dust extinction $A_{UV}$ in magnitudes from the UV continuum slope $\beta$. This fit is described as
\begin{equation}
    A_{UV} = 3.06 + 1.58\beta,
    \label{eq:Auvbeta}
\end{equation}
and has a significant scatter, which has been ignored here. Furthermore, Eq. (\ref{eq:Auvbeta}) can result in negative values of $A_{UV}$ which are set to zero. Although, this relationship has been observed in local galaxies, we are using it for dust correction in galaxies between $3\leq z \leq 8$. From this, we obtain a value of $A_{UV}$ for each $M_{UV}$ bin, and it is seen that dust affects brighter galaxies more. However, it must be noted that many galaxies could be fainter due to heavy dust obscuration, which makes these results biased against such an obscured population. Correction for dust can be included by replacing $M$ with $M-A_{UV}(M)$ in Eq (\ref{eq:magdensity}), except in $\phi(M)$.

\section{Observable Correlations in Baseband data} \label{app:corr_baseband}

\begin{figure*}
    \gridline{\fig{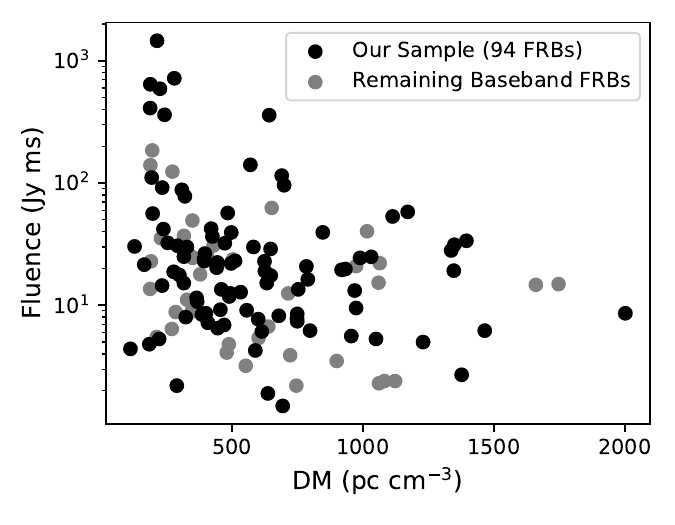}{0.49\textwidth}{(a)}
    \fig{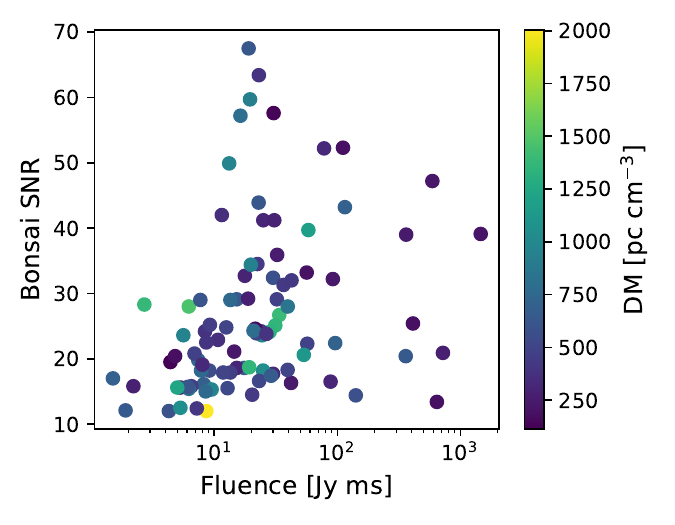}{0.49\textwidth}{(b)}
    }
    % \epsscale{1.0}
    % \plottwo{fluence_DM_scatter.pdf}{fluence_SNR_scatter.pdf}
    \caption{(a) Scatter plot of DM and fluence of baseband bursts, with black points being in our sample and gray points not. The empty space on the top-right shows the trend that bursts with high DM do not exhibit high fluences. (b) Scatter plot of fluence and $\rm bonsai$ SNR of the 94 baseband bursts in our sample, with the color of each point representing the DM. While low fluence bursts typically have low SNR, even high fluence bursts can have low SNR.}
    \label{fig:scatter_dm_fluence}
\end{figure*}

The CHIME/FRB Baseband Catalog 1 provides accurate fluence information, not available in the intensity catalog. Figure \ref{fig:scatter_dm_fluence}(a) shows the correlation between the fluence and DM, whose basis is outlined in Section 3. Figure \ref{fig:scatter_dm_fluence}(b) visually shows that the scatter in an SNR-fluence plot is significant, as has been discussed in Section \ref{sec:Edist_and_rate}.

% \begin{figure*}
%     \centering
%     \includegraphics[width=0.5\linewidth]{fluence_DM_scatter.pdf}
%     \caption{Scatter plot of DM and fluence of baseband bursts, with black points being in our sample and gray points not. The empty space on the top-right shows the trend that bursts with high DM do not exhibit high fluences. The apparent absence of very low fluence bursts (especially with high DM) is a selection effect, not a trend.}
%     \label{fig:scatter_dm_fluence}
% \end{figure*}

\section{Estimation of the Dispersion Measure Distribution from the Intergalactic Medium} \label{sec:appb}
The dispersion measures contributed by the IGM vary significantly depending on sightline and redshift of the source. Unlike the mapping of DM$_{\rm MW, ISM}$ with pulsars, mapping DM$_{\rm IGM}$ with any accuracy is extremely difficult, if not impossible. Useful estimates can now be extracted from large-scale simulations probing the cosmic electron density, but even they stumble at providing good small-scale resolution at the level of galaxies. Nevertheless, we utilize the work of \citet{Jaroszynski2019} to parametrize the DM$_{\rm IGM}$ distribution.

To generate a continuous family of distributions from the moments of the simulation-obtained DM-IGM distribution, we use the 4-parameter family of sinh-arcsinh distributions proposed by \citet{Jones2009}. The four parameters, namely, mode $m$, scale $s$, $\epsilon$ describing the skew, and $\delta$ parametrizing the kurtosis, are able to generate both symmetric and skewed distributions starting from a normal distribution. The reason for choosing such a distribution is that at low redshifts the DM-IGM distribution can be quite skewed, but it becomes increasingly symmetric at higher redshifts.

\begin{figure*}
    \centering
    \includegraphics[width=0.5\linewidth]{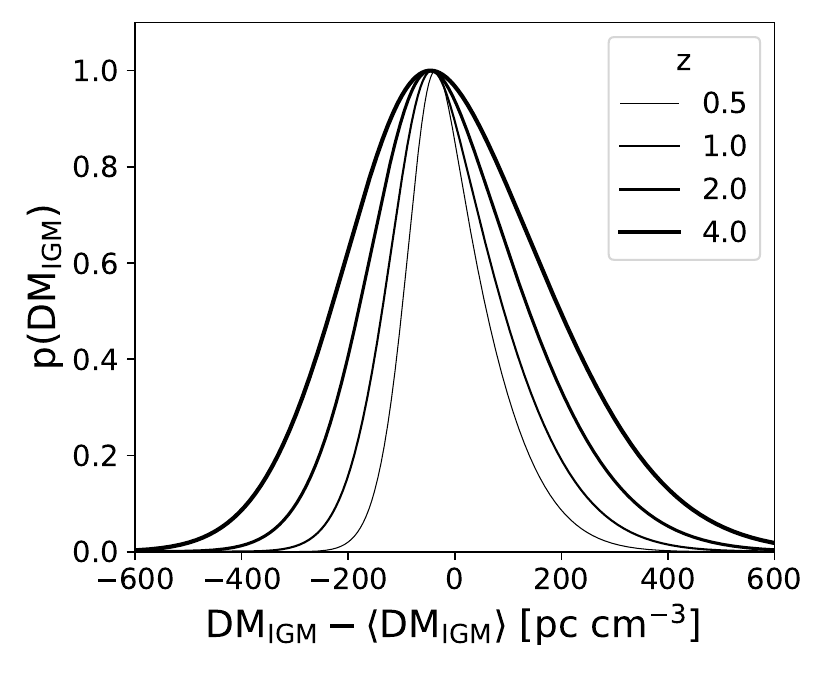}
    \caption{The normalized $\rm DM_{IGM}$ distribution at various redshifts whose mean is given by Eq. (\ref{eq:dmigm}) and whose 3 higher order moments have been calibrated with information in \citet{Jaroszynski2019}.}
    \label{fig:jaro_dmigm_fits}
\end{figure*}

Using the skewness and kurtosis of the DM-IGM distributions measured in \citet{Jaroszynski2019} corresponding to different redshifts, the values of $\epsilon$ and $\delta$ that yield the same skewness and kurtosis to within some tolerance are found. Since all DM-IGM distributions will be positively skewed, we consider $\epsilon > 0$. Furthermore, as redshift decreases, the distribution is expected to become increasingly skewed. We fit $\epsilon$ with a power-law function, and obtain the fitted form as,
\begin{equation}
    \epsilon = -0.0796 \times {\rm DM}^{0.204} + 0.5996.
    \label{eq:epsilon}
\end{equation}
Since $\delta < 1$ for a right-skewed distribution that mimics the low-redshift properties, and is equal to 1 for a Gaussian distribution, we fit an exponential function for $\delta$ as,
\begin{equation}
    \delta = 1 - 0.1903 e^{-0.0004 \times{\rm DM}}.
    \label{eq:delta}
\end{equation}
Similarly, the scale is fitted as a power law,
\begin{equation}
    s = 2.8526 \times {\rm DM}^{0.5059}.
\end{equation}
All the above fitting formulae accept the mode of the distribution, which is different from its average. Since we obtain the average DM-IGM from Eq (\ref{eq:dmigm}), we compute the relationship between the mode and the mean of the distribution for a range of modal values on a logarithmic scale between $10^{-1} - 10^4$ pc cm$^{-3}$. Consequently, for any $\langle {\rm DM_{ IGM}} \rangle$, we can estimate the nearest value of mode and generate the distribution with the same 2nd, 3rd and 4th moments as in \citet{Jaroszynski2019}. The resultant distributions are shown in Figure \ref{fig:jaro_dmigm_fits}, and their shape can be directly compared with Figure 3 of \citet{Jaroszynski2019}. It is important to mention that $\rm \langle DM_{IGM} \rangle$ computed in our work is different from \citet{Jaroszynski2019}.

\section{Estimating the survey duration of CHIME baseband system} \label{sec:base_survey_duration}
The baseband system was deployed and data was collected alongside Catalog 1 between 2018 December 9 and 2019 July 1 \citep{BasebandCatalog1_2023}. This system started operating four months after the beginning of Catalog 1, and the in the first few months of operations, baseband data for several FRBs was partially or completely lost. To estimate the survey duration of the baseband system, we need to perform an empirical analysis comparing the operational efficiency of the baseband system with the CHIME/FRB pipeline.

The survey duration of the CHIME/FRB pipeline $\Delta t = 214.8$ days \citep{CHIME2021}, which is used on the FRBs surviving selection cuts to estimate the all-sky rate. Excluding the pre-commissioning period, the telescope was operating from 2018 August 28 to 2019 July 1, giving an operational time of $\Delta t_{\rm op} = 308$ days. Instead of assuming that the telescope was operating optimally for the entirety of $\Delta t_{\rm op}$, which is likely not a correct assumption to make, it is implicitly assumed that the ratio of survey duration and operational time is the same for both Catalog 1 post-commissioning and the baseband system. Without getting into system-specific details about extracting exact down-time information for both systems, it is just easier to compare the ratio of the number of high-quality bursts recorded by both systems to obtain a rough estimate of the ``baseband survey duration'' $\Delta t_{\rm base}$.

The operational time of the baseband system $\Delta t_{\rm base, op} = 205$ days, and CHIME/FRB detected 202 high-quality bursts during this period. With 94 FRBs, we estimate that the baseband pipeline has an average efficiency $\eta_{\rm base} = 0.46$ of recording the bursts that are in the selection-cut sub-sample of Catalog 1. A simple calculation yields the baseband survey duration
\begin{equation}
    \Delta t_{\rm base} = \eta_{\rm base} \Delta t_{\rm base, op} \frac{\Delta t}{\Delta t_{\rm op}} = 65.8 \; \rm days.
\end{equation}

\section{Implementation of additional delay-time distributions} \label{sec:delay_time}
We implement two delay time distributions that are described by a single parameter, in addition to $\alpha$, $\gamma$ and \echar. The first one is called the ``delayed-tau'' model, and its probability distribution function is given as,
\begin{equation}
    P(t) = \frac{t}{\tau^2} {\rm exp}(-t/\tau).
\end{equation}
This distribution is sometimes used to parametrically model the star formation history of individual galaxies; but it is not able to replicate the cosmic star formation history for an ensemble of observed galaxies \citep{Carnall2019}. It is noted to have an extended tail. In contrast, the second distribution implemented here, called the Rayleigh distribution, does not have an extended tail. Its density function is given as,
\begin{equation}
    P(t) = \frac{t}{\tau^2} {\rm exp}\Big(\frac{-t^2}{2\tau^2}\Big).
\end{equation}

\begin{figure*}
    \epsscale{1.1}
    \plottwo{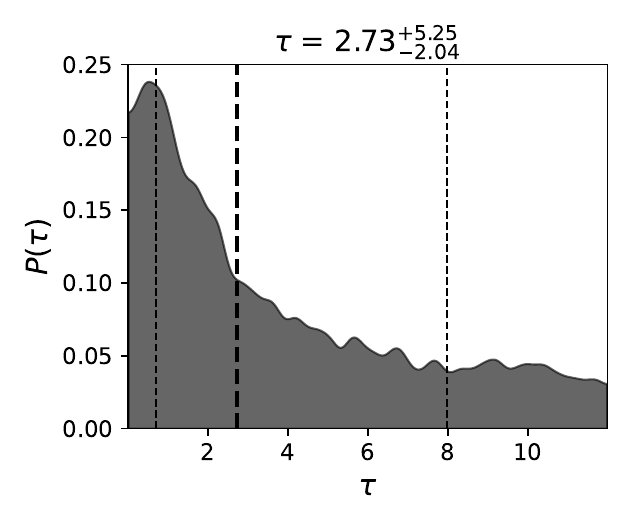}{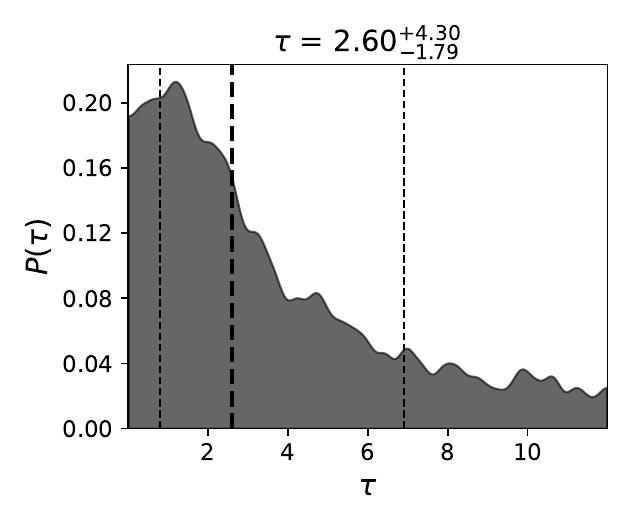}
    \caption{Posterior distributions of the mode $\tau$ describing (\textit{left}) the delayed-tau distribution, and (\textit{right}) the Rayleigh distribution. Other parameters are very similar to the CDT model parameters.
    }
    \label{fig:cornerplot_param_delaydist}
\end{figure*}

Implementing MCMC gives a hint of unresolved peaks in both delay time posteriors, following which nested-sampling is used \citep{dynesty:2020}. The posteriors and medians for $\alpha$, $\gamma$ and \echar are consistent between MCMC and nested-sampling, and additionally, are very similar to those obtained for the CDT model. The posteriors for $\tau$, which describe the mode of both models, are shown in Figure \ref{fig:cornerplot_param_delaydist}. For a given mode, the median of the delayed-tau distribution is larger than that of the Rayleigh distribution. That is why the most likely values at $\approx 0.7$ Gyr and at $\approx 1.2$ Gyr for the delayed-tau and the Rayleigh distributions, respectively, correspond to a median delay time of roughly $1.3$ Gyr. It is, in turn, consistent with the range of likely values of $\tau$ predicted by the CDT model
Unlike the CDT model, both these models do not have negligible posterior probabilities at characteristic delay times $\tau \gtrsim 6$ Gyr because a fraction of their PDFs cover $0 < t < 6$ Gyr delays as well. 

\section{Effect of pulse width on the redshift distribution of Fast Radio Bursts} \label{sec:app_pulse_width}

FRB pulses are cosmologically dilated when they originate at high redshift, in addition to the instrumental effect of intra-channel dispersion smearing. As such, the pulse arriving at Earth would have a width of $w_{\rm arr} = w_{\rm int}(1+z)$, and the simultaneous dispersion smearing term is $w_{\rm DM} = 8.3\times 10^{-3} \;{\rm ms \; DM_{pc/cm^3}} \Delta \nu_{\rm MHz} \nu_{\rm GHz}^{-3}$ \citep{Cordes2003}. $w_{\rm int}$ is the intrinsic temporal width of the FRB, and for CHIME, $\Delta \nu_{\rm MHz} = 0.0244$ is the frequency channel width of the detector in MHz, and $\nu_{0,\rm GHz}=0.6$ is the observing frequency in GHz. Then, $w_{\rm DM} = 9.4 \times 10^{-4} \;{\rm ms \; DM_{pc/cm^3}}$, with ${\rm DM_{pc/cm^3}}$ being the total DM given by Eq (\ref{eq:dm_components}). We approximate $\rm DM_{MW, ISM} + DM_{MW, halo} = 100$ pc cm$^{-3}$, and for a given z, we take the mean $\rm DM_{IGM}(z)$ (Eq \ref{eq:dmigm}) and the mean $\rm DM_{host}$ for an SMD-tracking population. Finally, the scintillation decorrelation bandwidth of bursts in their rest frame $\Delta \nu_{\rm dc} \propto [\nu_0(1+z)]^{4.4}$, which being a very strong function of redshift, implies that scattering contributed by host galaxies $w_{\rm scat} \propto (\Delta \nu_{\rm dc})^{-1}$ becomes negligible for increasing redshifts. Thus, the observed pulse for an FRB at high redshift has an effective width of \citep{LorimerKramer2012, Gardenier2019},
\begin{equation}
    w_{\rm eff} = \sqrt{w_{\rm arr}^2 + w_{\rm DM}^2 + w_{\rm scat}^2 + w_{\rm samp}^2} \gtrsim \sqrt{w_{\rm int}^2(1+z)^2 + w_{\rm DM}^2 + w_{\rm samp}^2}
    \label{eq:app_pulse_width}
\end{equation}
Here, $w_{\rm samp}=0.983$ ms $ \sim 1$ ms is the sampling time of CHIME. Not considering the scattering timescale gives us a crude lower limit on the effective pulse width.

Using \citet{Cordes2003}, we can define the detectability criteria as ${\rm SNR^{th}} = F_{\nu_0}^{\rm th}(z)/\sqrt{w_{\rm eff}}$. The sensitivity threshold for a telescope is generally defined as ${\rm SNR^{th}} = F_{\nu_0}^{\rm th}(z=0)/\sqrt{1 \; \rm ms}$. Thus, the threshold fluence can be expressed as 
\begin{equation}
    F_{\nu_0}^{\rm th}(z) = F_{\nu_0}^{\rm th}(z=0) \sqrt{\frac{w_{\rm eff}(z)}{1\; \rm ms}}.
\end{equation}
Because of the nature of Eq (\ref{eq:app_pulse_width}), we get a lower limit on $F_{\nu_0}^{\rm th}(z)$ assuming a burst with intrinsic duration of 1 ms. 

Since the sensitivity of CHIME has not been modelled as a function of the observed pulse width along with DM and fluence, it is difficult to analyze it on the same footing as the other two observables. 
Based on their selection analysis, \citet{CHIME2021} conclude that there is little evidence for an intrinsically wide population at low redshifts, because that is where CHIME is limited to observing. They further show that CHIME seems to be missing only a small fraction of bursts between $1-10$ ms. 
% It is hard to prove, but can be assumed that the slight decrease at higher pulse widths may be due to pulse broadening of observed FRBs that form the higher redshift subset of the full sample.
Even without any compensation, CHIME may not be missing a noticeable fraction of bursts based on their pulse width. Consequently, a calibration of the data with a model of just DM and fluence would remain valid within the CHIME observable redshift range, thereby allowing the recreation of the observed redshift distribution of bursts accurately without invoking pulse width broadening.

As it is shown above that CHIME can observe roughly the full range of intrinsic pulse widths for local Universe FRBs, it may not be wrong to assume that approximately the same range of pulse widths from FRBs at redshift $z$ is visible above the scaled threshold fluence $F_{\nu_0}^{\rm th}(z)$. Error from the omission of the scattering term is present predominantly either when the Galactic scattering is present with low latitude FRBs, or when host galaxy scattering is present for low redshift FRBs. The intrinsic FRB redshift distribution for $F_{\nu_0}^{\rm th}(z=0) > 1$ Jy ms is plotted in Figure \ref{fig:bestfit_zEdist}(b) with pulse width correction included (black lines) and excluded (gray lines).


\bibliography{sample631}{}
\bibliographystyle{aasjournal}



\end{document}